\documentclass{aa}  

\usepackage{graphicx}
\usepackage{txfonts}
\usepackage{psfig}
\usepackage{amsmath}
\usepackage{amssymb}
\usepackage{upgreek}
\usepackage{longtable}
\usepackage{lscape}
\usepackage[mathscr]{eucal}
\usepackage{float}
\usepackage{afterpage}
\usepackage{subcaption}
\usepackage[export]{adjustbox}
\usepackage[usenames, dvipsnames]{color}
\usepackage{titlesec}
\usepackage{caption}
\usepackage{booktabs}

\begin{document}

   \title{Expansion Kinematics of Young Clusters. I. Lambda Ori}


   \author{Joseph J. Armstrong
          \inst{1}
          \and
          Jonathan C. Tan\inst{1}
          }

   \institute{Department of Space, Earth \& Environment, Chalmers University of Technology, SE-412 96 Gothenburg, Sweden\\
              \email{joseph.armstrong@chalmers.se}
             }

   \date{Received July 16, 2024}

 
  \abstract
   {Most stars form in clusters or associations, but only a small number of these groups are expected to remain bound for longer than a few megayears. Once star formation has ended and the molecular gas around young stellar objects has been expelled via feedback processes, most initially bound young clusters lose the majority of their binding mass and begin to disperse into the Galactic field.}
   {This process can be investigated by analysing the structure and kinematic trends in nearby young clusters, particularly by analyzing the trend of expansion, which is a tell-tale sign that a cluster is no longer gravitationally bound and dispersing into the field.}
   {We combined \textit{Gaia} DR3 five-parameter astrometry with calibrated RVs for members of the nearby young cluster $\lambda$ Ori (Collinder 69).}
   {We characterised the plane-of-sky substructure of the cluster using the $Q$-parameter and the angular dispersion parameter. We find evidence that the cluster contains a significant substructure but that this is preferentially located away from the central cluster core, which is smooth and likely remains bound. We found strong evidence for expansion in $\lambda$ Ori in the plane of sky by using a number of metrics, but we also found that the trends are asymmetric at the 5$\sigma$ significance level, with the maximum rate of expansion being directed nearly parallel to the Galactic plane. We subsequently inverted the maximum rate of expansion of $0.144^{+0.003}_{-0.003}$ kms$^{-1}$pc$^{-1}$ to give an expansion timescale of $6.944^{+0.148}_{-0.142}\:$Myr, which is slightly larger than the typical literature age estimates for the cluster. We also found asymmetry in the velocity dispersion as well as signatures of cluster rotation, and we calculated the kinematic ages for individual cluster members by tracing their motion back in time to their closest approach to the cluster centre.}
   {}


   \maketitle
%

\section{Introduction}

Most stars form in clusters \citep{lada03, gutermuth09} or within substructures of larger associations and star-forming regions \citep{wright20}. The clusters form in giant molecular clouds (GMCs) when the collapse and fragmentation of protocluster gas clumps produce many young stars in a small volume, which are bound together by their mutual gravity and that of the surrounding gas in which they are embedded. As a consequence, the initial structure and dynamics of groups of young stars are expected to reflect that of their parent molecular cloud. That is, they may be turbulent and clumpy \citep{kuhn14,sills18} and potentially retain kinematic signatures of any large-scale compressive motions that may have triggered star formation, such as from galactic shear driven GMC-GMC collisions \citep[e.g.][]{tan00} or collisions driven by stellar feedback \citep[e.g.][]{inutsuka15}. 

However, once massive stars form in an embedded cluster, their feedback is often expected to expel the gas and dust in the vicinity, and the cluster loses most of its binding mass. Hence, many young stars become unbound and begin to disperse into the field after residual gas expulsion \citep{kroupa01b,goodwin06}. It is unclear how many embedded clusters and substructures in star-forming regions will merge into bound open clusters that can survive this process. The kinematic signature of expansion in a group of young stars, often characterised by trends in velocity-position space in a given direction, is a tell-tale sign that the group is dispersing.

The dynamical evolution of the cluster can also lead to the dispersal of its member stars, either by the ejection of cluster members as runaway or walkaway stars via strong or moderate dynamical interactions \citep{farias20,kounkel22} or by shearing due to the gravity of nearby molecular clouds or Galactic tidal forces. The latter of these can produce tidal tails over hundreds of megayears \citep[e.g,][]{kroupa22}. 

Kinematic studies of star clusters and star-forming regions have been revolutionised by the availability of high-precision five-parameter astrometry from \textit{Gaia}, which provides positions, proper motions, and parallaxes (and thus distances) for nearly two billion sources \citep{Gaiaedr3}. It is now possible to identify hundreds of low-mass cluster members by characterising overdensities in both positional and proper motion space with the use of clustering tools such as HDBSCAN without the need for complementary age indicators. However, \textit{Gaia} lacks radial velocity (RV) information for sources fainter than Gmag = 14, that is, the vast majority of low-mass cluster members, and thus to obtain complete 3D velocity information, it is necessary to combine \textit{Gaia} astrometry with spectroscopic RVs from surveys such as APOGEE, GALAH, and \textit{Gaia}-ESO. With this precise kinematic information, we can investigate in great detail the expansion trends of young clusters.

\citet{kuhn19} studied the plane-of-sky kinematics of a sample of young clusters observed in the MYSTIX program, where young cluster members were identified by their enhanced X-ray activity. In particular, for a majority of the clusters in their sample (75\%), they found evidence of expansion, indicating that a significant fraction of young stars in these regions are unbound and dispersing into the Galactic field. \citet{guilherme23} used an integrated nested Laplace approximation (INLA) to reconstruct the velocity fields of open clusters and reported expansion patterns for 14 clusters and rotation patterns for eight. \citet{wright19} found evidence of strongly anisotropic expansion in the young cluster NGC 6530 directed preferentially in the direction of declination (Dec). 

Expansion trends have also been reported in many OB associations and star-forming regions, such as Orion OB1 \citep{kounkel18,zari19}; Vela OB2 \citep{cantatgaudin19b,armstrong20,Armstrong22}; and Sco-Cen \citep{wright18}, which are sparse and substructured and thus expected to be dispersing. But again, these expansion trends were found to be significantly anisotropic in many cases, bringing into doubt the classical interpretation of these groups having originated as relaxed compact clusters since expansion after residual gas expulsion in this case is expected to be isotropic. Instead, the kinematic evidence points to OB associations forming in large volumes with an initial substructure over timescales of many megayears. However, whether the anisotropy of their expansion is inherited from the turbulent motion of the parent molecular cloud or the large-scale flows associated with the triggering of the star-forming clump or is a consequence of tidal forces (e.g. induced by a nearby GMC) is not yet understood. 

The $\lambda$ Ori cluster is a nearby ($\sim400\:$pc) young \citep[$\sim4-6$ Myr;][]{zari19,cao22} cluster to the north of the Orion complex. \citet{kounkel18} found evidence for radial expansion in the plane of sky and attributed it to a supernova explosion $\sim4.8$ Myr ago, which is also responsible for dissipating most of the molecular gas in the vicinity. \citet{zari19} then estimated a kinematic age of 8 Myr for $\lambda$ Ori based on the expansion trends seen in their sample identified using the DBSCAN algorithm whilst also finding a best-fitting isochronal age of $5.6^{+0.4}_{-0.1}$ Myr, indicating that $\lambda$ Ori likely formed with substantial initial volume and substructure. \citet{cao22} used the SPOTS stellar evolution models in conjunction with individual extinction values for cluster members with spectra from the APOGEE survey to estimate a best-fitting isochronal age of $3.9\pm0.2$ Myr as well as an intrinsic age spread of $\sim0.35$ dex. 

We combined astrometry from the latest \textit{Gaia} data release (DR3) with calibrated RVs \citep{tsantaki22} for members of $\lambda$ Ori identified in the \textit{Gaia} catalogue \citep{cantat-gaudin20}. We filtered sources based on their cluster membership probabilities and the quality of their astrometry in order to obtain a clean sample for our kinematic study. We analysed the cluster substructure on the plane of sky using the Q-parameter and angular dispersion parameter \citep[ADP;][]{dario14}. We looked for signatures of expansion and calculated the cluster expansion velocity, the expansion timescale, and the asymmetry of the expansion, as well as the velocity dispersions and rotation trends. We also performed plane-of-sky traceback for individual cluster members to the point of closest approach to the cluster centre and compared these results to the ages estimated by comparison to stellar evolution models. We discuss our results in light of previous studies and give an overview of the past dynamical evolution of the cluster.

\begin{figure}
\begin{center}
\includegraphics[width=\columnwidth]{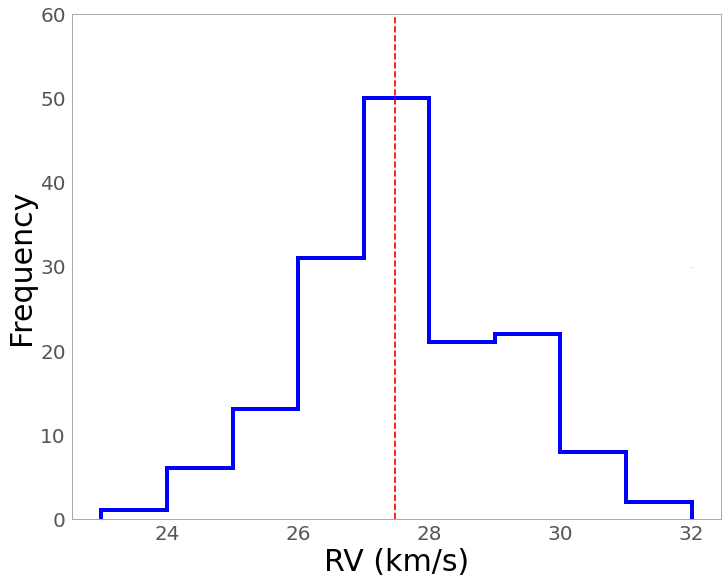}
\caption{Histogram of the RVs for 189 members of $\lambda$ Ori that match with \protect\citet{tsantaki22}. The median cluster RV of 27.47 km$s^{-1}$ is indicated by the red dashed line.}
\label{RVs}
\end{center}
\end{figure}

\begin{figure*}
\begin{center}
\includegraphics[width=500pt]{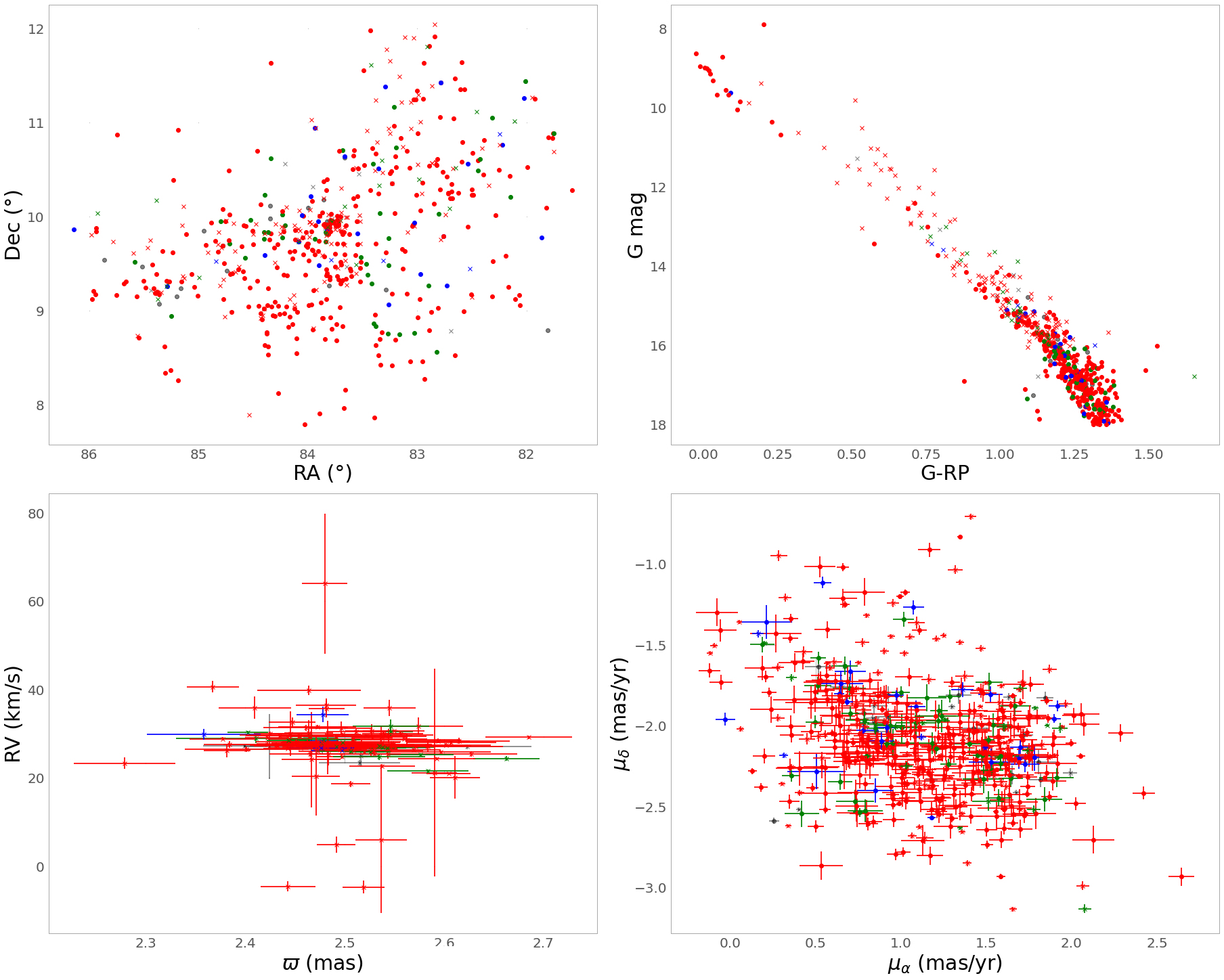}
\caption{Gaia data of cluster members . \textit{(a) Top left:} Sky positions of cluster members of $\lambda$ Ori from \protect\citet{cantat-gaudin20}. Here, sources with a membership probability of $1.0$ are plotted in red, $0.9$ in green, $0.8$ in blue, and $0.7$ in grey.
Sources with RVs from \protect\citet{tsantaki22} are plotted as crosses. \textit{(b) Top right:} \textit{Gaia} EDR3 BP-RP versus Gmag colour-magnitude diagram. \textit{(c) Bottom left:} \textit{Gaia} EDR3 parallax versus RV. \textit{(d) Bottom right:} \textit{Gaia} EDR3 proper motions of the sources.}%
\label{ClusterPlots}
\end{center}
\end{figure*}

\section{Data}
\label{data}
We began with the open cluster catalogue of \citet{cantat-gaudin20}, which provides mean positions, proper motions, distances, and ages for 2017 open clusters, as well as the list of individual members from the \textit{Gaia} DR2 catalogue and their cluster membership probabilities. The cluster parameters are based on $\geq$0.7 probability cluster members selected using the UPMASK method \citep{krone-martins14} applied to data from \textit{Gaia} DR2. For the cluster $\lambda$ Ori (Collinder 69) there are 604 total candidate members in the catalogue. In order to have a clean sample of cluster members with high precision astrometry we select the $\lambda$ Ori cluster members and match them with \textit{Gaia} DR3, which has improved astrometric precision over DR2, and then filter out sources with \textit{Gaia} DR3 RUWE$>1.4$ which is the suggested threshold for astrometric quality \citep{Gaiaedr3}.
This gives a sample of 563 high probability members of the $\lambda$ Ori cluster for further kinematic analysis.

We calculated the mean sky positions in right ascension (RA), Dec, and $l$, $b$ -- and proper motions from our sample of high-probability members, and we estimated the uncertainties on these values with a bootstrapping approach, calculating these means and medians from 100,000 randomly selected (with replacement) samples of cluster members and taking the 84th and 16th percentiles of the posterior distributions as the uncertainties. These values are given in Table~\ref{kinematic_table}. Using the same approach with the parallaxes of the high-probability members, we also find a median distance for the cluster of $399.68^{+0.89}_{-0.89}$ pc.

\subsection{Radial velocities}
We then matched these cluster members to the {Survey of Surveys} \citep{tsantaki22} compilation of RVs, which combines RVs from large-scale spectroscopic surveys including \textit{Gaia}, APOGEE, GALAH, \textit{Gaia}-ESO, RAVE, and LAMOST into a single cross-calibrated sample. When combining RV information from multiple surveys, care must be taken to account for differences in instruments used, the selection functions of observed targets and the analysis methods used to derive RVs, all of which contribute the heterogeneity of the survey samples. \citet{tsantaki22} find for example that RVs from the LAMOST survey are offset by 5.18 km$^{-1}$ from RVs from \textit{Gaia}.   

We find matches with 189 of the $\lambda$ Ori cluster members (Fig. \ref{RVs}), of which 164 have RVs from APOGEE, 29 from LAMOST and 32 from \textit{Gaia}. This is sufficient to calculate an average RV and its dispersion for the cluster and to correct for projection effects (see Sect. \ref{Tangential velocities}), but not for an analysis of the full 3D kinematic trends. The median RV we calculate is $27.47^{+0.08}_{-0.08}$ kms$^{-1}$ and the observed RV dispersion is $2.14$ kms$^{-1}$.

\subsection{Overlap with other samples}
We find that 182 cluster members match with the sample analysed in \citet{cao22} and 64 with \citet{dolan01}. Also, 336 cluster members match with the \textit{Gaia} DR3 catalogue of YSOs identified by variability \citep{marton23}.

\section{Structure and morphology}
\label{structure}

\subsection{$Q$ parameter}
\label{Qparametersection}

\begin{figure} 
    \subfloat{{\includegraphics[width=\columnwidth]{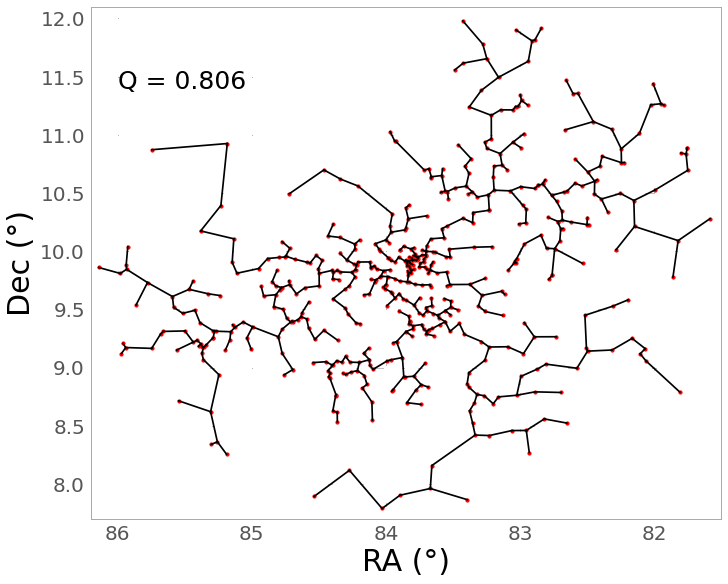} }}%
    \setlength{\belowcaptionskip}{-10pt}
    \setlength{\textfloatsep}{0pt}
    \caption{
    Right ascension and Dec positions of $\lambda$ Ori cluster members from \protect\citet{cantat-gaudin20} (red). Minimum spanning tree branches are shown in black.} 
    \label{Qparameter}%
\end{figure}

One commonly used measure of substructure in star-forming regions is the minimum spanning tree $Q$ parameter \citep{cartwright04}. The $Q$ parameter is defined as the ratio between the mean edge length of the minimum spanning tree $\Bar{m}$ and the mean edge length of the complete graph $\Bar{s}$. Typically, a value of $Q<0.7$ indicates that the region is relatively clumpy and substructured, while $Q>0.9$ indicates that the region is smoother and more centrally concentrated \citep{parker22}. 

We apply this method to the high-probability members of $\lambda$ Ori and normalise $\Bar{m}$ to $\frac{\sqrt{NA}}{N-1}$, where $N$ is the number of cluster members and $A$ is the area of the smallest circle with radius $R$ which encompasses them. $\Bar{s}$ is normalised to $R$.

For the RA, Dec positions of members of $\lambda$ Ori we obtain Q $= 0.806$ (Fig. \ref{Qparameter} \textit{top}), putting the cluster in the neither obviously substructured nor smooth category. This would indicate that some amount of initial substructure still remains in the current configuration of the cluster, but it has begun to be erased. This is similar to $Q$ values derived for other nearby young clusters, such as the 22 clusters analysed by \citet{jaehnig15}, the $Q$ values for which were all between 0.74 and 0.89 when considering members within each cluster's half-mass radius.

\subsection{The centre of $\lambda$ Ori}
\label{center}

\begin{figure}
\begin{center}
\includegraphics[width=\columnwidth]{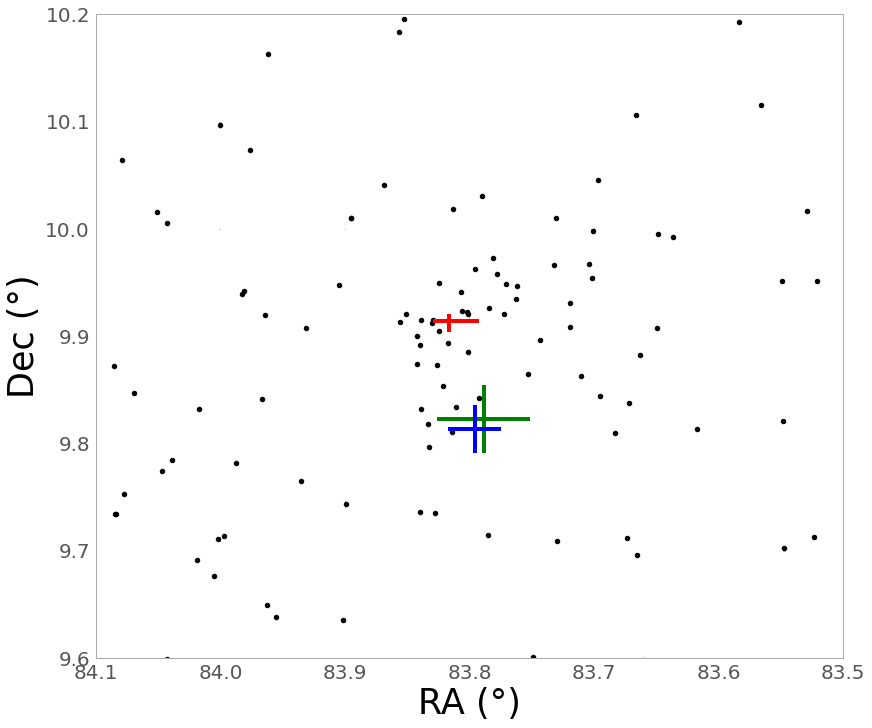}
\caption{Right ascension and Dec positions of high-probability $\lambda$ Ori cluster members from \protect\citet{cantat-gaudin20} (black). The mean position (green), median position (blue), and centre of mass (red) are shown, and we have also plotted their respective uncertainties.}
\label{centers}
\end{center}
\end{figure}

We consider the centre of the cluster to be at its centre of mass rather than a geometric centre. In order to determine this we follow the iterative approach introduced and applied to the ONC by \citet{dario14}, which consists of determining the centre of mass for all cluster members contained within iteratively smaller apertures, each centred on the previous aperture's centre of mass. We use the same approach for the high-probability members of $\lambda$ Ori, assuming equal masses, reducing the aperture radius by 20\% each iteration and limiting the minimum aperture size to a 3 arcmin radius. We estimate the uncertainties of the centre of mass with a bootstrapping approach similar to that used for the mean and median positions (Sect. \ref{data}) sampled from the posterior distribution of 100,000 iterations of the whole procedure. We find the centre of mass to be located at $l={195.074^{+0.018}_{-0.018}}^\circ$, $b={-11.981^{+0.008}_{-0.009}}^\circ$, which we take as the cluster centre in the following analyses. This position is noticeably offset from the mean and median positions by $>0.1^\circ$ in Dec. The position of the derived centre of mass in RA and Dec is shown relative to the mean and median positions of high-probability cluster members in Fig.~\ref{centers} with their associated errors.

\subsection{Two-dimensional ellipse fitting}
\label{ellipse}

\begin{figure}
\begin{center}
\includegraphics[width=\columnwidth]{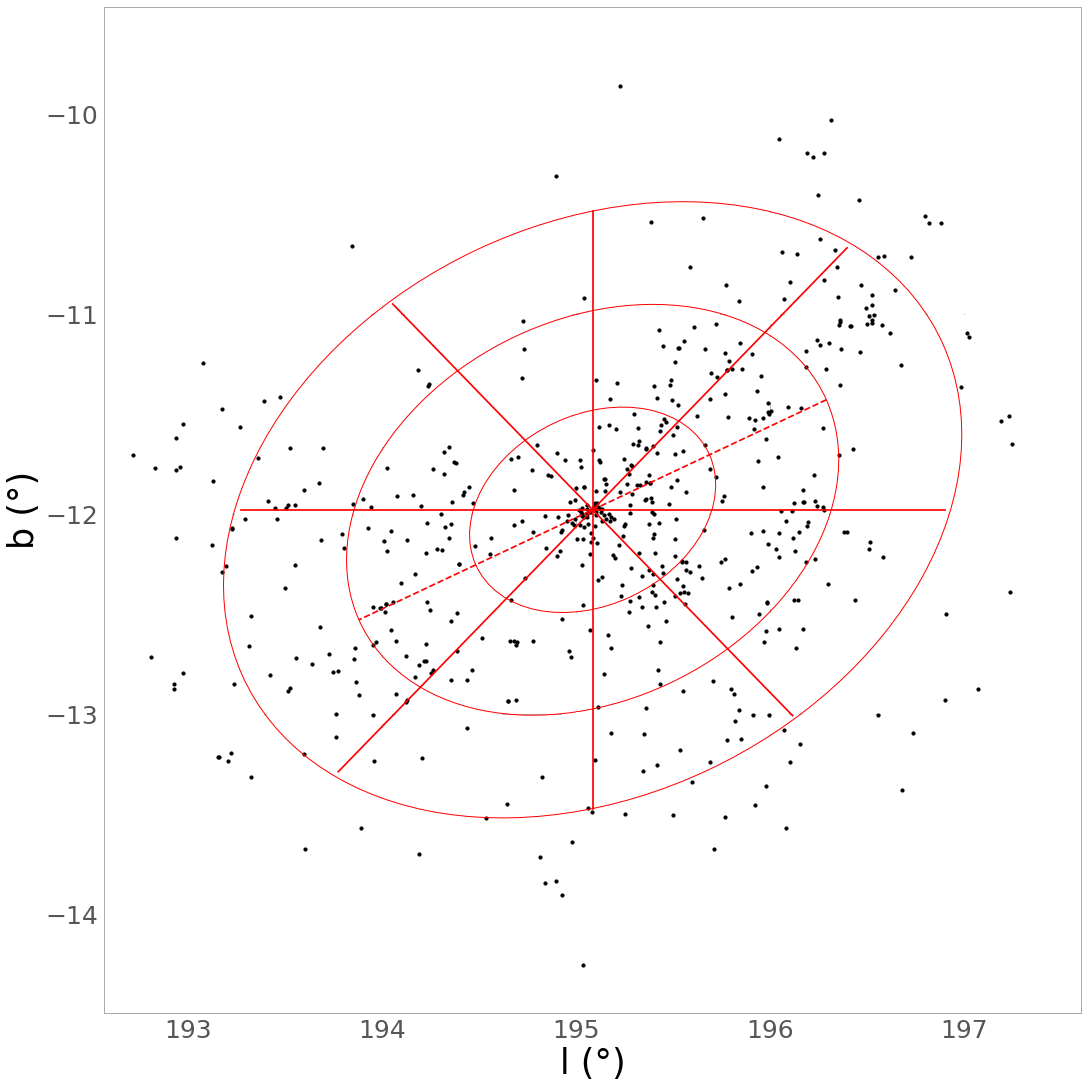}
\caption{Galactic longitude and latitude of members of $\lambda$ Ori. The ellipses are centred on the estimated 2D centre of mass (see \ref{center}) and fitted to the cluster member positions by least squares (see \ref{ellipse}). The dashed line indicates the semi-major axis of the best-fit ellipse, which has an eccentricity of $0.695$ and an orientation of the semi-major axis at $24.3^\circ$ counterclockwise to the direction of positive Galactic longitude. The solid lines indicate the division of the elliptical annuli into different sectors, which are used to calculate the ADP ($\delta_{\rm ADP,e,N}$). }
\label{Ellipse2D}
\end{center}
\end{figure}

\begin{figure*}
\begin{center}
\includegraphics[width=\textwidth]{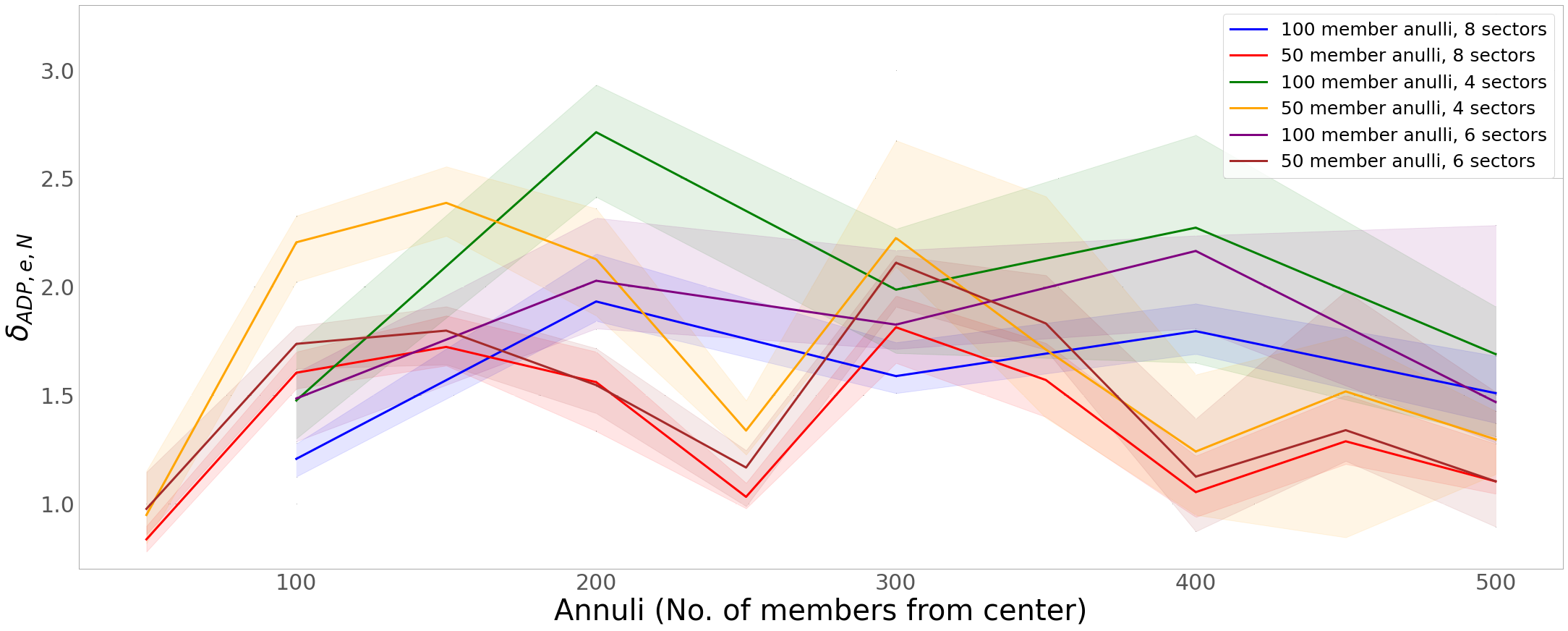}
\caption{Angular dispersion parameter ($\delta_{ADP,e,N}$) with eight sectors per concentric annuli containing 50 (red) or 100 (blue) members, six sectors per concentric annuli containing 50 (brown) or 100 (purple) members, and four sectors with 50 (yellow) or 100 (green) members. The $\delta_{ADP,e,N}(r)$ values are calculated for orientations of sectors rotated 1$^\circ$ at a time, and we plot the 50th (solid lines), 16th, and 84th percentile values (shaded regions) for each annulus. }
\label{ADP}
\end{center}
\end{figure*}

We fit an ellipse to the Galactic sky coordinates of members of $\lambda$ Ori using least squares minimisation of the edge of the ellipse to the data points (skimage.ellipsemodel) given the central coordinates determined in Sect.~\ref{center}.
We find the best-fitting ellipse to have an eccentricity of $e=0.695$ and with the semi-major axis oriented at $24.3^\circ$ counterclockwise to the direction of positive Galactic longitude (see Fig. \ref{Ellipse2D}). 

\begin{figure}
        \includegraphics[width=250pt]{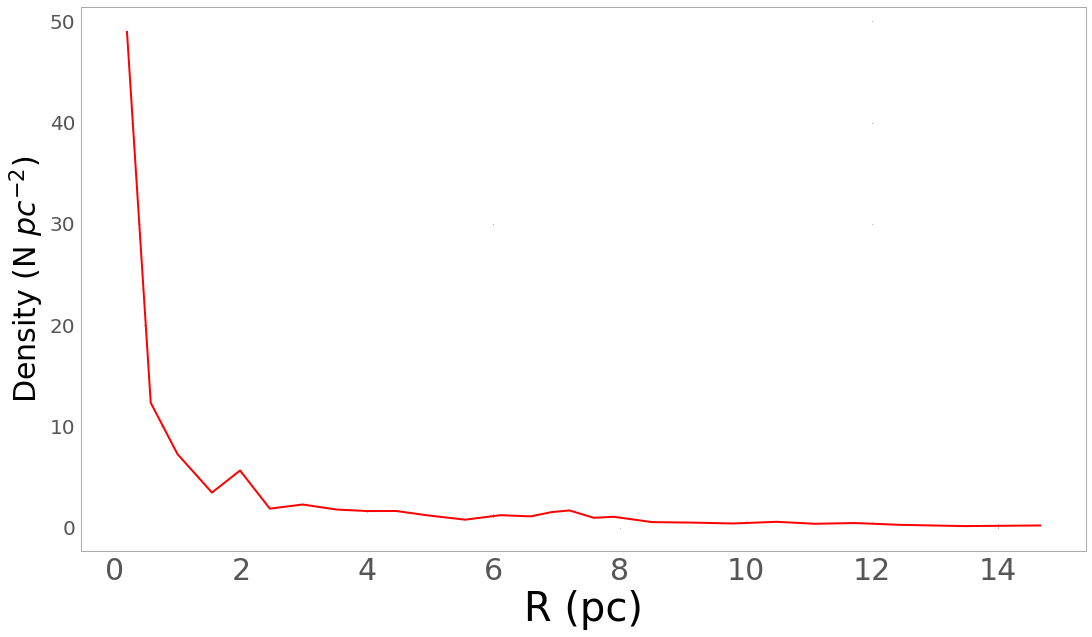}
        \setlength{\belowcaptionskip}{-10pt}
        \setlength{\textfloatsep}{0pt}
        \caption{Density profile of the 563 high probability members of the $\lambda$ Ori cluster from \protect\citet{cantat-gaudin20}. Cluster members are binned by increasing radial distance from the cluster centre $R$ (pc) in bins of 20 members each. For each bin, the average $R$ is plotted against the area density of that bin (N pc$^{-2}$) where $N=20$ and the area is defined as the area of the smallest circle (centred on the cluster center) containing all the members of a bin, minus the area of previous bins. }
        \label{Densityprofile}
\end{figure}

\begin{figure*}
        \includegraphics[width=500pt]{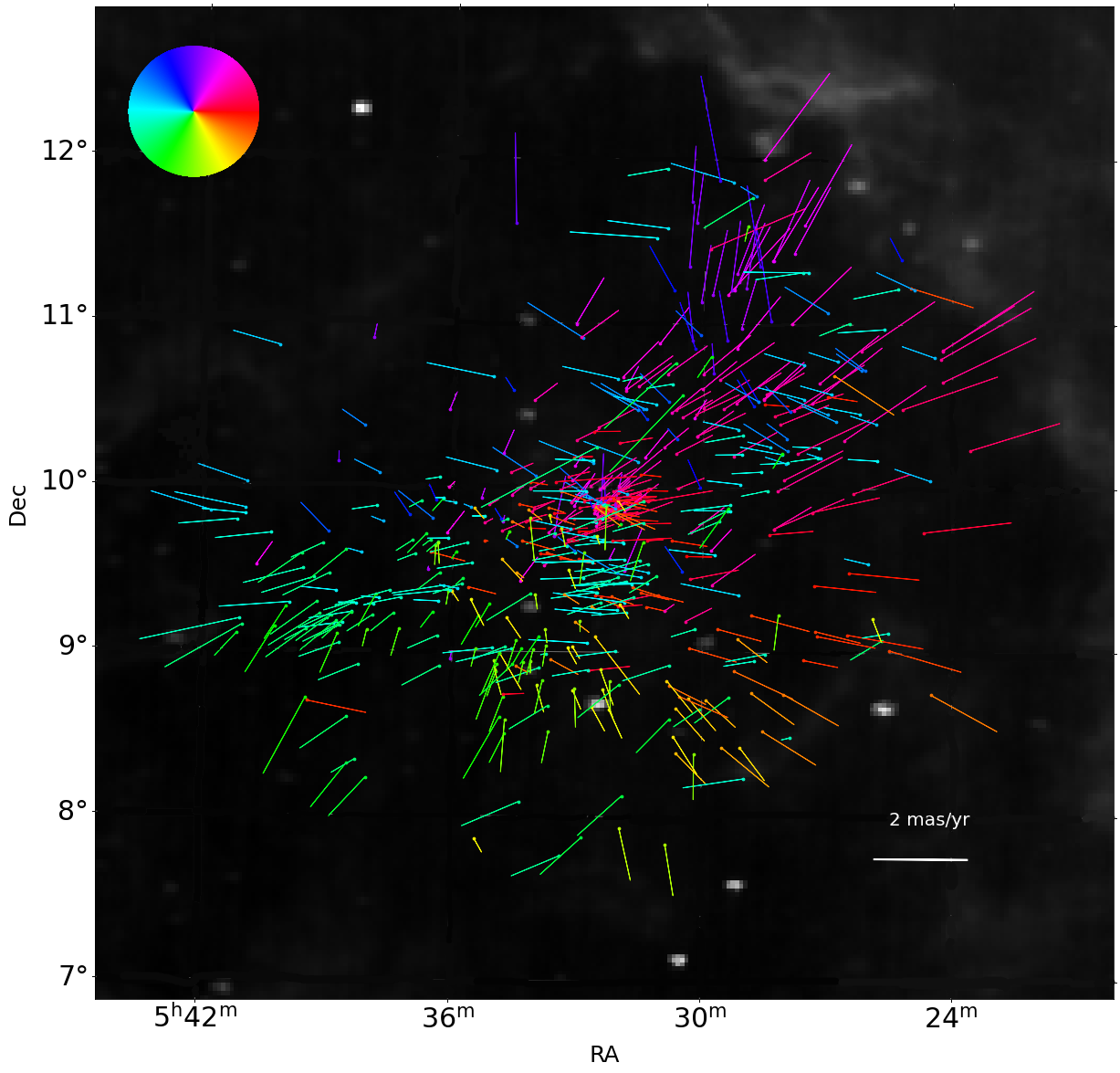}
        \setlength{\belowcaptionskip}{-10pt}
        \setlength{\textfloatsep}{0pt}
        \caption{Spatial distribution of the 563 high probability members of the $\lambda$ Ori cluster from \protect\citet{cantat-gaudin20}. The vectors indicate the proper motion relative to the cluster. Points are colour-coded based on the position angle of the proper motion vector (see the colour wheel in the top left as a key). The magnitude scale (masyr$^{-1}$) of proper motion vectors is indicated by the scale bar in the bottom right.}
        \label{map}
\end{figure*}

\begin{figure}
\begin{subfigure}{0.49\textwidth}
    \includegraphics[width=245pt]{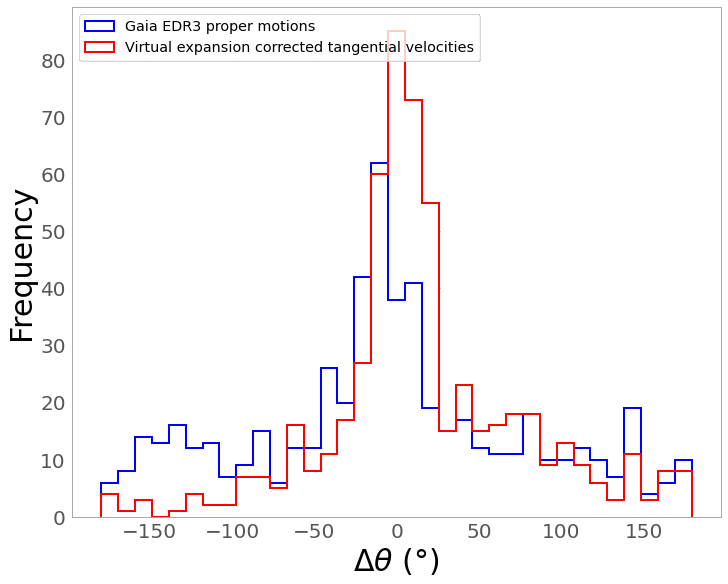}
\end{subfigure}
\begin{subfigure}{0.49\textwidth}
    \includegraphics[width=\columnwidth]{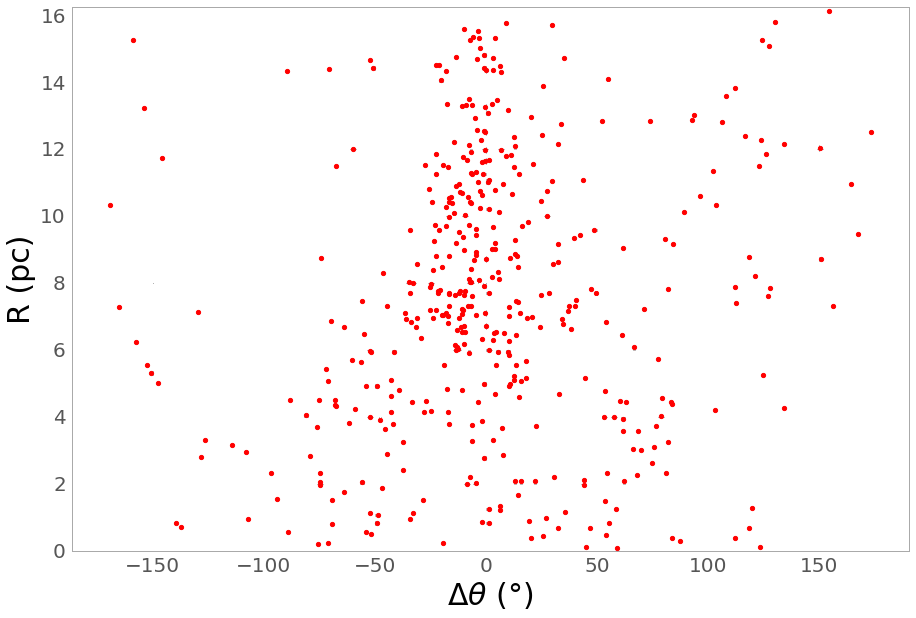}
\end{subfigure}
\begin{subfigure}{0.49\textwidth}
    \includegraphics[width=\columnwidth]{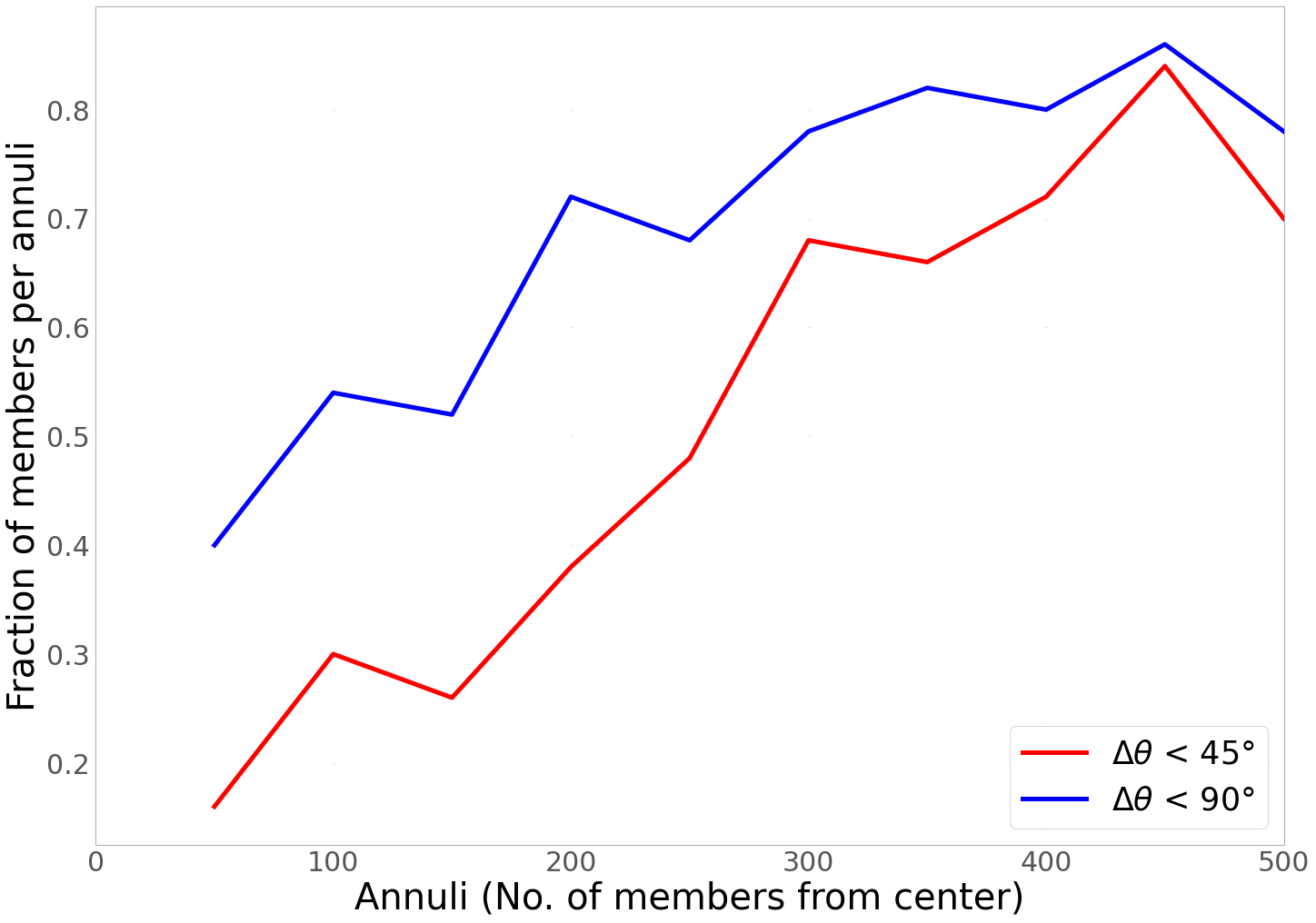}
\end{subfigure}
\caption{Velocity angle - position angle ($\Delta\theta$) trends. \textit{Top:} Histogram of difference in angle between proper motion vector (blue), virtual-expansion corrected tangential velocity vector (red), and cluster member position relative to the cluster centre for members of $\lambda$ Ori ($\Delta\theta$). \textit{Middle:} Difference in angle between virtual-expansion corrected tangential velocity vector and cluster member position relative to the cluster centre ($\Delta\theta$) against radial distance ($R$). \textit{Bottom:} Fraction of members with virtual-expansion corrected tangential velocities directed within 45$^\circ$ of the direction towards to the cluster centre (red) and within 90$^\circ$ of the direction towards to the cluster centre (blue) per concentric annuli containing 50 members. }
\label{Vangle}
\end{figure}

\subsection{Angular dispersion parameter}
\label{AngularDispersionParameter}

Using the above ellipse as the basis we also investigate the angular substructure of the $\lambda$ Ori cluster using the ADP, $\delta_{\rm ADP,e,N}(r)$, as introduced by \citet{dario14}. $\delta_{\rm ADP,e,N}(r)$ is calculated by dividing the cluster into a number of concentric elliptical annuli (of ellipticity $e$), which are then further divided into $N$ sectors (see example in Fig.~\ref{Ellipse2D}), and counting the number of cluster members located in each sector of each annulus. $\delta_{\rm ADP,e,N}$ then corresponds to the standard deviation of counts per sector of a given annulus, and $\delta_{\rm ADP,e,N}(r)$ to the variation of $\delta_{\rm ADP,e,N}$ in annuli of increasing radial distance. As \citet{dario14} demonstrate, low ($<1$) values of $\delta_{\rm ADP,e,N}$ indicate a very smooth, likely dynamically old distribution, such as that found in Globular clusters, while high ($>2.5$) values indicate a highly substructured, dynamically young distribution similar to what has been found in the Taurus star-forming region. The young but somewhat dynamically evolved ONC was found to have $\delta_{\rm ADP,e,N}$ values between $1.3 - 1.8$ for variations of annulus size and number of sectors. \citet{jaehnig15} also calculated $\delta_{\rm ADP,e,N}(r)$ for a sample of 22 young clusters characterised in the MYStIX survey and found typical values between $1 - 2$, with the overall trend that $\delta_{\rm ADP,e,N}$ values tend to increase the more cluster members further from the cluster centre are included. Young clusters may be expected to exhibit smoother distributions at their centers if these are gravitationally bound and with relatively high stellar density, since a greater density of cluster members will mean a higher rate of stellar interactions, speeding up the time required to erase any initial substructure.

\citet{dario14} explain that, in order for the ADP $\delta_{\rm ADP,e,N}(r)$ to be compared between different clusters the number of cluster members per annulus must be fixed. We therefore calculated $\delta_{\rm ADP,e,8}$, $\delta_{\rm ADP,e,6}$ and $\delta_{\rm ADP,e,4}$ for concentric annuli containing 50 and 100 high-probability members of $\lambda$ Ori, and, in order to account for the deviation of the measured parameter $\delta_{\rm ADP,e,N}(r)$ due to the orientation of the sectors, we calculate each for orientations of sectors rotated 1$^\circ$ at a time. We plot the 50th percentile values of the resulting distribution of $\delta_{\rm ADP,e,N}$ values (solid lines) and take as their respective uncertainties as the 16th and 84th percentile values (shaded regions) as shown in Fig.~\ref{ADP}. 

While there is a noticeable difference in the trends of $\delta_{\rm ADP,e,N}(r)$ for different size annuli, the trends for same size annuli with different numbers of sectors are very similar. $\delta_{\rm ADP,e,N}$ values for same size anulli overall are larger for fewer sectors, but this difference shrinks as the distance from the cluster centre increases. 

For $\delta_{\rm ADP,e,N}(r)$ trends with 100 cluster members per annulus $\delta_{\rm ADP,e,N}$ increases from the cluster centre to a peak at the third annulus (containing the 201st - 300th cluster member), which is similar to trends seen in clusters in \citet{jaehnig15}, but then decreases in the fourth annulus, indicating that there is significant substructure up to $\sim1^\circ$ away from the cluster center, but the outskirts of the cluster beyond this are relatively smooth. 

For annuli with 50 members the $\delta_{\rm ADP,e,N}(r)$ trends exhibit many features the 100 member annuli $\delta_{\rm ADP,e,N}(r)$ do not. There is a large difference in $\delta_{\rm ADP,e,N}$ between the centre ($\sim1$, very smooth) and the 50th - 100th member annulus ($2 - 3$, highly substructured), which is averaged out in the 100 member annulus. The peak of high $\delta_{\rm ADP,e,N}$ at the 200th - 250th member annulus likely indicates the same substructure as the peak of the 100 member annuli $\delta_{\rm ADP,e,N}(r)$ does, but is made more distinct by the low $\delta_{\rm ADP,e,N}$ values in the annuli either side of it. 

The mean $\delta_{\rm ADP,e,8}$ for 50 member annuli (red in Fig.~\ref{ADP}) is 1.412, and 1.701 for 100 member annuli (blue in Fig.~\ref{ADP}). The mean $\delta_{\rm ADP,e,6}$ for 50 member annuli (brown in Fig.~\ref{ADP}) is 1.539, and 1.885 for 100 member annuli (purple in Fig.~\ref{ADP}). The mean $\delta_{\rm ADP,e,4}$ for 50 member annuli (yellow in Fig.~\ref{ADP}) is 1.824, and 2.116 for 100 member annuli (green in Fig.~\ref{ADP}). These mean values are similar to those calculated for the ONC by \citet{dario14}, albeit slightly higher for $\delta_{\rm ADP,e,4}$, indicating that $\lambda$ Ori is more substructured than the ONC, which has likely undergone significant dynamical evolution, but considerably less than a sparse, low-mass star-forming region such as Taurus.

It is worth considering how the cluster selection method employed by \citet{cantat-gaudin20} may have affected these results. While their approach, identifying clusters as dense groups in astrometric parameter space, is successful in selecting cluster members with similar astrometry to the cluster average with high confidence, it is likely to miss cluster members with motions distinct from the cluster. Particularly, for unbound clusters dispersing into the field, cluster members are less likely to be identified the further they are from the cluster centre. Thus, the tailing-off of $\delta_{ADP,e,N}$ values in the outer annuli we see in $\lambda$ Ori may be due in part to incompleteness of cluster membership in the outskirts of the cluster.

\subsection{Density profile}
\label{densityprofile}

We investigate the radial density profile of the $\lambda$ Ori cluster, choosing not to account for ellipticity for the sake of simplicity. Cluster members are binned by increasing radial distance from the cluster centre $R$ (pc) in bins of 20 members each. For each bin, the area density of that bin (N pc$^{-2}$), where $N=20$, is defined as the area of the smallest circle (centred on the cluster center) containing all the members of a bin, minus the areas of interior bins. The resulting radial density profile is shown in Fig.~\ref{Densityprofile}.

Unlike studies of other clusters \citep[e.g.][]{hillenbrand98} where there is a flattening of density in the central region that we don't find for $\lambda$ Ori, we do not define the core radius $r_c$ (pc) by fitting a King profile \citep{king62}. Instead, we adopt the same approach as \citet{tarricq22}, where we take $r_c$ as the radius at which the density is half the peak density. This yields a core radius of $r_c = 0.44$ (pc).

We find that 25 cluster members are located within this core radius. We fit an ellipse to the coordinates of these members using the same approach as Sect.~\ref{ellipse}, and the same central coordinates, to look for evidence of elongation within the core. We find the best-fitting ellipse to have an eccentricity of $e = 0.791$ with the semi-major axis oriented at $6.3^{\circ}$ counterclockwise to the direction of positive Galactic longitude, similar to the results found for the best-fitting ellipse for all cluster members. 

Nine of these cluster core members have RVs in \citet{tsantaki22}, originating from APOGEE and \textit{Gaia}. They occupy a large range, from $25.82$ kms$^{-1}$ to $30.66$ kms$^{-1}$, though binarity is a possible contributor to this. A correlation between these RVs and position on the sky would indicate a component of cluster core rotation in the line-of-sight, but we find no such correlation for these nine members with RVs. 

\section{Expansion}
\label{Expansion}

If a cluster or association of young stars becomes gravitationally unbound, its member stars will tend to move away from each other, and so the group expands. In the following section we present and compare several methods of identifying and measuring expansion rates, timescales and asymmetry, and the results obtained by applying them to our high-quality astrometry sample of members of $\lambda$ Ori. 

\subsection{Proper motions}
\label{Proper motions}
In order to study the internal kinematics of the cluster we first need to transform the observed proper motions of cluster members into the reference frame of the cluster. The cluster mean proper motions are $\mu_{\alpha_0} = 1.137^{+0.020}_{-0.020}$ (masyr$^{-1}$), $\mu_{\delta_0} = -2.079^{+0.015}_{-0.014}$ (masyr$^{-1}$) (Sect.~\ref{data}). 

Figure~\ref{map} shows the RA, Dec positions of the members of the $\lambda$ Ori cluster with vectors indicating their proper motions relative to the cluster mean, colour-coded based on the position angle of the vector. In general, it can be seen that the majority of cluster members are moving outwards from the cluster center, especially those on the outskirts, indicating that the cluster is expanding as a whole.

Rather than take the mean proper motions of all 563 cluster members as the central cluster velocity we instead take the mean proper motions of cluster members belonging to the dense cluster core, as defined in Sect.~\ref{densityprofile}, which are are $\mu_{\alpha_0} = 0.734^{+0.033}_{-0.030}$ (masyr$^{-1}$), $\mu_{\delta_0} = -2.011^{+0.018}_{-0.019}$ (masyr$^{-1}$).

\subsection{Tangential velocities}
\label{Tangential velocities}
 
However, one must take into account a number of effects which may confound evidence for expansion derived from uncorrected proper motions. A cluster's line-of-sight motion, either approaching or receding from the observer, will produce components of proper motion of its members often known as 'virtual expansion' \citep{cantatgaudin19b}. If RVs for a sufficient number of cluster members are available the cluster's mean RV can be used to correct this effect. 

We transform proper motions into tangential velocities, using the cluster's median distance and including a correction for 'virtual expansion' using the cluster's median RV following the equations of \citet{brown97}, using a Bayesian approach. We sample the posterior distribution function using the MCMC sampler emcee \citep{emcee}. We perform 1000 iterations with 100 walkers per star in an unconstrained parameter space with the only prior being that the line-of-sight distance of the star must be $<10$ kpc. We discard the first half of our iterations as a burn in and we report the medians of the posterior distribution function as the best fit, with the 16th and 84th percentiles as the 1$\sigma$ uncertainties.

The central tangential velocities of the cluster core are $v_{l_0}  = 4.046^{+0.079}_{-0.077}$ (kms$^{-1}$) and  $v_{b_0} = -0.669^{+0.045}_{-0.049}$ (kms$^{-1}$).

 \begin{figure*}
\sidecaption
  \includegraphics[width=12cm]{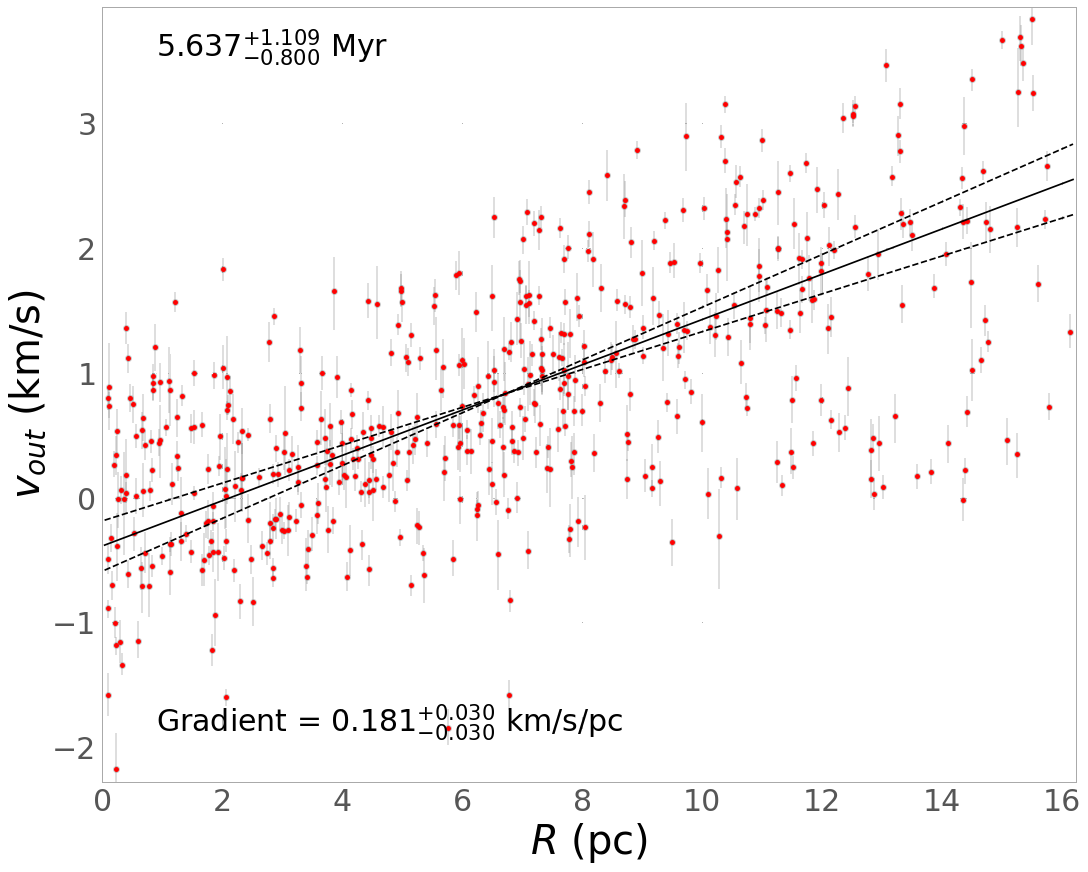}
     \caption{Distance from cluster centre (pc) versus $v_{out}$ (kms$^{-1}$). The uncertainties for members of $\lambda$ Ori are in red. The best-fitting gradients and the 16th and 84th percentiles values of the fit are shown as solid and dashed lines respectively in each panel. The best-fit gradient (rate of expansion) and uncertainties are given in kms$^{-1}$pc$^{-1}$ as well as the corresponding timescale of expansion and uncertainties in Myrs. }
     \label{1D}
\end{figure*}

\subsection{Expansion velocity}
\label{Expansion velocity}

\citet{kuhn19} measured expansion velocities $v_{\rm out}$, the velocity component of cluster members directed radially from the cluster center, and argue that a significantly positive median expansion velocity for a cluster is indicative of global expansion. They found that 75$\%$ of clusters in their sample exhibited significantly positive expansion velocities.

In order to isolate the expansion velocity we separate the tangential velocities of cluster members into radial and transverse components, that is, components directed radially away from the cluster centre and perpendicular to the cluster center, which can be interpreted as expansion velocity $v_{\rm out}$ and rotational velocity $v_{\rm rot}$ components. 

Here we measure the cluster expansion velocity by taking the median expansion velocity component of all cluster members for 1,000,000 iterations with additional uncertainties randomly sampled from the observed velocity uncertainties. The uncertainties on the cluster expansion velocity are then taken as the 16th and 84th percentile values of the posterior distribution. 

We find a median expansion velocity of $0.711^{+0.021}_{-0.021}$ km$s^{-1}$ for $\lambda$ Ori, which is positive at the $>$30$\sigma$ significance level, strong evidence that the cluster is expanding. This is a very reasonable expansion velocity for a young cluster, especially considering the large range of expansion velocities calculated by \citet{kuhn19}, which range from $2.07\pm1.10$ kms$^{-1}$ for $G353.1+0.6$ to $-2.06\pm1.00$ kms$^{-1}$ for M17. The expansion velocity we find for $\lambda$ Ori is most closely comparable to those found for IC5146 ($0.48\pm0.25$ km~s$^{-1}$) and V454 Cep ($0.55\pm0.34$ km~s$^{-1}$). 

We note that the median expansion velocity obtained using velocities without correcting for 'virtual expansion' is $0.337^{+0.021}_{-0.021}$ kms$^{-1}$, significantly less than the corrected result, highlighting the importance of this correction when attempting to measure expansion trends in the plane of sky. Expansion analyses using uncorrected proper motions cannot give reliable expansion velocities, rates or timescales, and there is a significant risk of false-positive and false-negative detections and non-detections of expansion.

\subsection{Velocity-position angle}
\label{Velocity-Position Angle}
Another method of identifying cluster-wide expansion is to investigate the distribution of directions in which cluster members are moving relative to the cluster centre. This is done by calculating the difference in angle between a cluster member's position relative to the cluster centre and the angle of orientation of it's relative velocity, which we thus denote as $\Delta\theta$ ($^\circ$). If a cluster member is moving directly away from the cluster centre it will have $\Delta\theta = 0^\circ$; thus the distribution of $\Delta\theta$ for cluster members of an unbound, dispersing cluster should show a peak close to $0^\circ$.

However, it is worth noting that a significant cluster-wide rotation trend would cause this peak to deviate from $0^\circ$, within the range $|\Delta\theta| < 90^\circ$. A peak value either $\Delta\theta < -90^\circ$ or $\Delta\theta > 90^\circ$ would indicate cluster-wide contraction.

In Fig.~\ref{Vangle}a we show histograms of the velocity-position angles ($\Delta\theta$) for members of $\lambda$ Ori relative to the cluster centre using uncorrected \textit{Gaia} EDR3 proper motions (blue) and virtual-expansion corrected tangential velocities (red). The typical uncertainty on this angle is 6.4$^{\circ}$. The clear peak around 0$^{\circ}$ shows that the majority of cluster members are moving directly outwards from the centre of the cluster. Indeed, 59$\%$ of the 550 high-probability members have tangential velocity-position angles $|\Delta\theta| < 45^\circ$ from the cluster center, and 74$\%$ $|\Delta\theta| < 90^\circ$, showing that the majority of cluster members are moving outwards from the center, evidence that this cluster is unbound and expanding from its initial configuration. 

The dispersion in $\Delta\theta$ indicates that the velocities of many cluster members have some component of rotation around the cluster center, with a slight preference to $\Delta\theta > 0^\circ$, suggesting a small trend of anti-clockwise rotation in the plane of sky.

We also investigate how the velocity-position angles of cluster members change with distance from the cluster centre. In Fig.~\ref{Vangle}b we plot $\Delta\theta$ against radial distance from the cluster centre $R$, and in Fig.~\ref{Vangle}c we show how the fraction of cluster members with $|\Delta\theta| < 45^\circ$ (red) and $|\Delta\theta| < 90^\circ$ (blue) per 50 member annuli of increasing $R$ (see Sect.~\ref{AngularDispersionParameter}). The overall trend apparent in both plots is that, while only a small fraction of cluster members close to the centre have motions consistent with expansion, this fraction increases with distance from the cluster centre $R$, up to $>80\%$ in the penultimate annulus, but drops slightly at the very outskirts. This may also be an effect of membership incompleteness in the cluster halo (see Sect.~\ref{AngularDispersionParameter}).

\subsection{One-dimensional position-velocity gradient}
\label{1Dexpansion}

In Fig.~\ref{1D} we plot distance from cluster centre $R$ (pc) against $v_{\rm out}$ (kms$^{-1}$) for members of $\lambda$ Ori. We determine the linear best-fit parameters using Bayesian inference and sample the posterior distribution function using the Markov chain Monte Carlo (MCMC) ensemble sampler \textit{emcee} \citep{emcee}, the same approach as described in \citet{Armstrong22}. We model the linear fit with parameters for the gradient, intersect and the fractional amount by which the uncertainties are underestimated ($m$, $b$, $f$). We assume that errors are Gaussian and independent, and estimate the maximum likelihood with linear least squares using the likelihood function from Eq. 6 of \citet{Armstrong22}. We account for uncertainties in position by varying the measured position according to its uncertainties during the MCMC simulation. This is repeated for 2000 iterations with 200 walkers, the first half of which are discarded as burn in. From the second half the medians and 16th and 84th percentiles are reported from the posterior distribution function as the linear best-fit gradient and uncertainties. The best-fit linear gradient in Fig.~\ref{1D} for $\lambda$ Ori is $0.181^{+0.030}_{-0.030}$ kms$^{-1}$pc$^{-1}$, evidence for cluster-wide expansion at the 6$\sigma$ significance level.

If a star cluster (or association) is expanding, and one assumes that the stars were originally in a more compact configuration then the expansion gradient in kms$^{-1}$pc$^{-1}$ can be inverted (with a unit conversion factor of 1.023) to estimate the timescale for the expansion in Myr, a type of kinematic age estimate. The corresponding timescale to the expansion rate determined here is $5.637^{+1.109}_{-0.800}$~Myr. 

One caveat to kinematic ages estimated from a 1D expansion trend is that they are an estimate of the time in the past when the cluster members would trace back to a point given their current velocities, which is an unphysical assumption. Thus, an expansion age should be considered an upper limit on the timescale for which a cluster has been expanding from its initial configuration.

Thus, if a cluster or association is assumed to have been formed initially unbound, one would expect to measure an expansion timescale in close agreement or slightly greater than an age estimated by comparison to stellar evolution models in colour-magnitude diagrams or Li-depletion, for example, while an expansion timescale lower than estimates from other age methods would suggest that some event subsequent to a cluster's formation has caused it to become unbound (such as residual gas expulsion).

In this case the expansion timescale of $5.637^{+1.109}_{-0.800}$~Myr is slightly smaller than the kinematic age of $\sim8$ Myr from \citet{zari19}, but in better agreement with the timescale of $\sim4.8$ Myr from \citet{kounkel18}. It is also in close agreement with the best-fitting isochronal age of $5.6^{+0.4}_{-0.1}$~Myr from \citet{zari19}, but greater than the best-fitting isochronal age of $3.9\pm0.2$~Myr from \citet{cao22}. The differences in these kinematic age estimates may be due to differences in the cluster membership samples used, while differences in the isochronal ages may also be due to the differences in stellar evolution models. However, in general, our expansion timescale is in agreement with or greater than recent isochronal age estimates, which would suggest that $\lambda$ Ori has been unbound and expanding since shortly after its formation. Still, the considerable scatter of points around the linear best-fit in Fig.~\ref{1D} also indicates that there is some variation in the rate of expansion among cluster members, which may well depend on the direction it is measured in. We seek to investigate this possible expansion asymmetry in the following analysis.

\subsection{Direction of maximal expansion}
\label{Direction of Maximal Expansion}

Since many expansion trends that have been detected in the literature are significantly anisotropic \citep[e.g.][]{wright19,Armstrong22}, we expect that such clusters will exhibit a maximum rate of expansion in a particular direction.
The direction of maximum expansion may help inform us as to which physical mechanisms have had the greatest impact on the disruption of a young cluster.

\begin{figure} 
    {\includegraphics[width=\columnwidth]{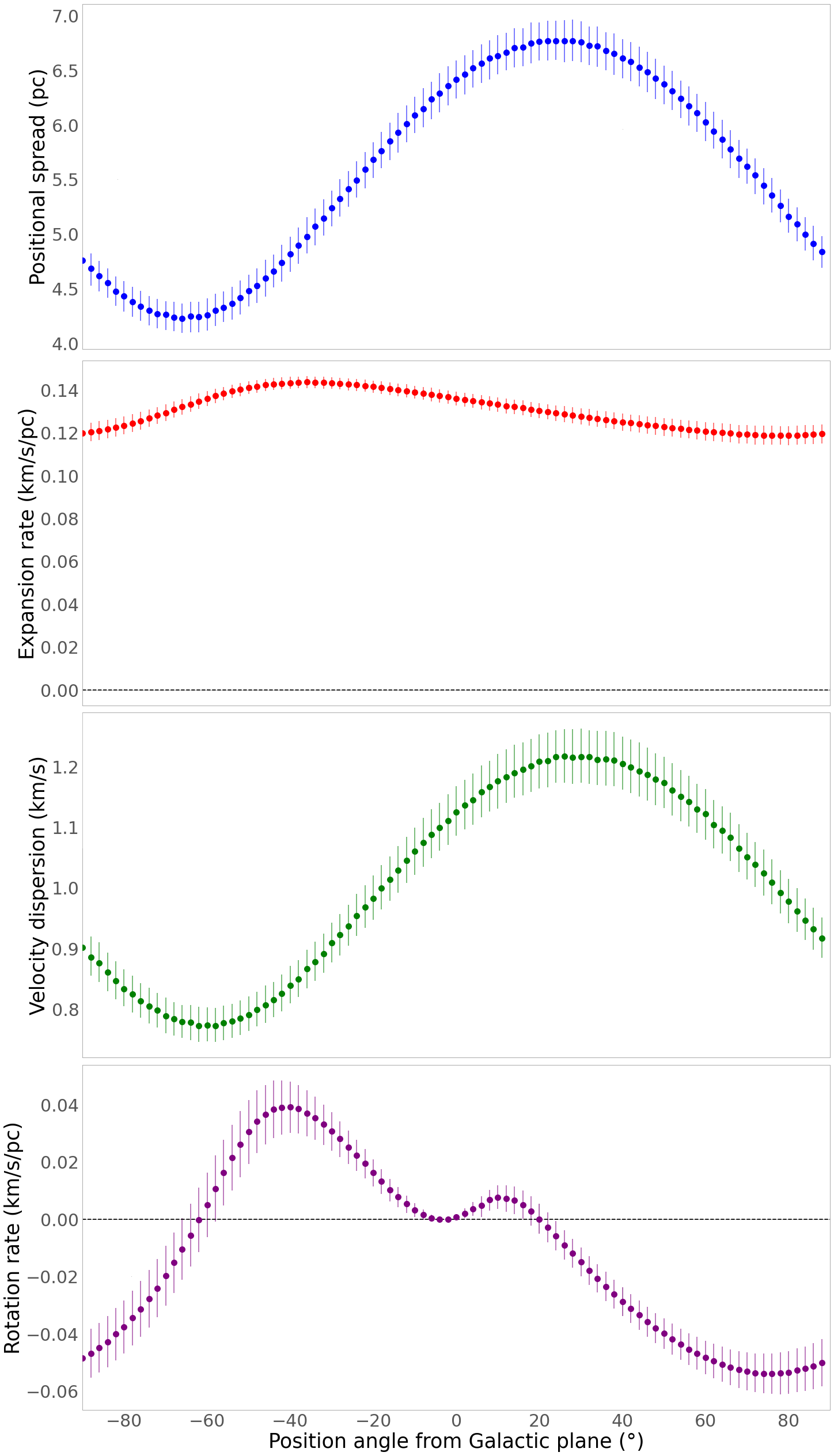} }%
    \setlength{\belowcaptionskip}{-10pt}
    \setlength{\textfloatsep}{0pt}
    \caption{Positional spread (blue), rate of expansion (red), 1D velocity dispersion (green), and rotation rate (purple) with uncertainties versus position angle from the Galactic plane for members of $\lambda$ Ori. The direction of maximum expansion is shown to be at -34$^{\circ}$ below the Galactic plane with increasing longitude while minimum expansion is in the direction of Galactic latitude. The rates of maximum and minimum expansion are different at the 5$\sigma$ confidence level, making the plane-of-sky expansion of $\lambda$ Ori significantly anisotropic. The 1D velocity dispersions are also anisotropic at the 8$\sigma$ confidence level, with the maximum dispersion of $1.225^{+0.048}_{-0.043}$ kms$^{-1}$ found at 32$^{\circ}$ above the Galactic plane with increasing longitude, close to the direction of maximum positional spread at 24$^{\circ}$ above the Galactic plane with increasing longitude. The minimum 1D velocity dispersion is $0.771^{+0.029}_{-0.026}$ kms$^{-1}$. The rotation rates are also anisotropic at the 8$\sigma$ confidence level, with the maximum rotation rate of $0.039^{+0.009}_{-0.009}$ kms$^{-1}$ found at -40$^{\circ}$ above the Galactic plane with increasing longitude, close to the direction of minimum positional spread at -64$^{\circ}$ above the Galactic plane with increasing longitude and the direction of maximum expansion. The minimum rotation rate is $-0.055^{+0.007}_{-0.007}$ kms$^{-1}$. }%
    \label{ExpansionAsymmetry}%
\end{figure}

In order to calculate the direction of maximum expansion we calculate the rate of expansion in one dimension ($l$ versus $v_{l}$) using MCMC as above (Sect.~\ref{1Dexpansion}), rotate the positions and velocities of cluster members by 2$^{\circ}$ and repeat until we have calculated rates of expansion for axes rotated by up to 180$^{\circ}$. We then plot the calculated rates of expansion and their uncertainties against the orientation of the axes (Fig.~\ref{ExpansionAsymmetry}).

\begin{figure} 
    {\includegraphics[width=\columnwidth]{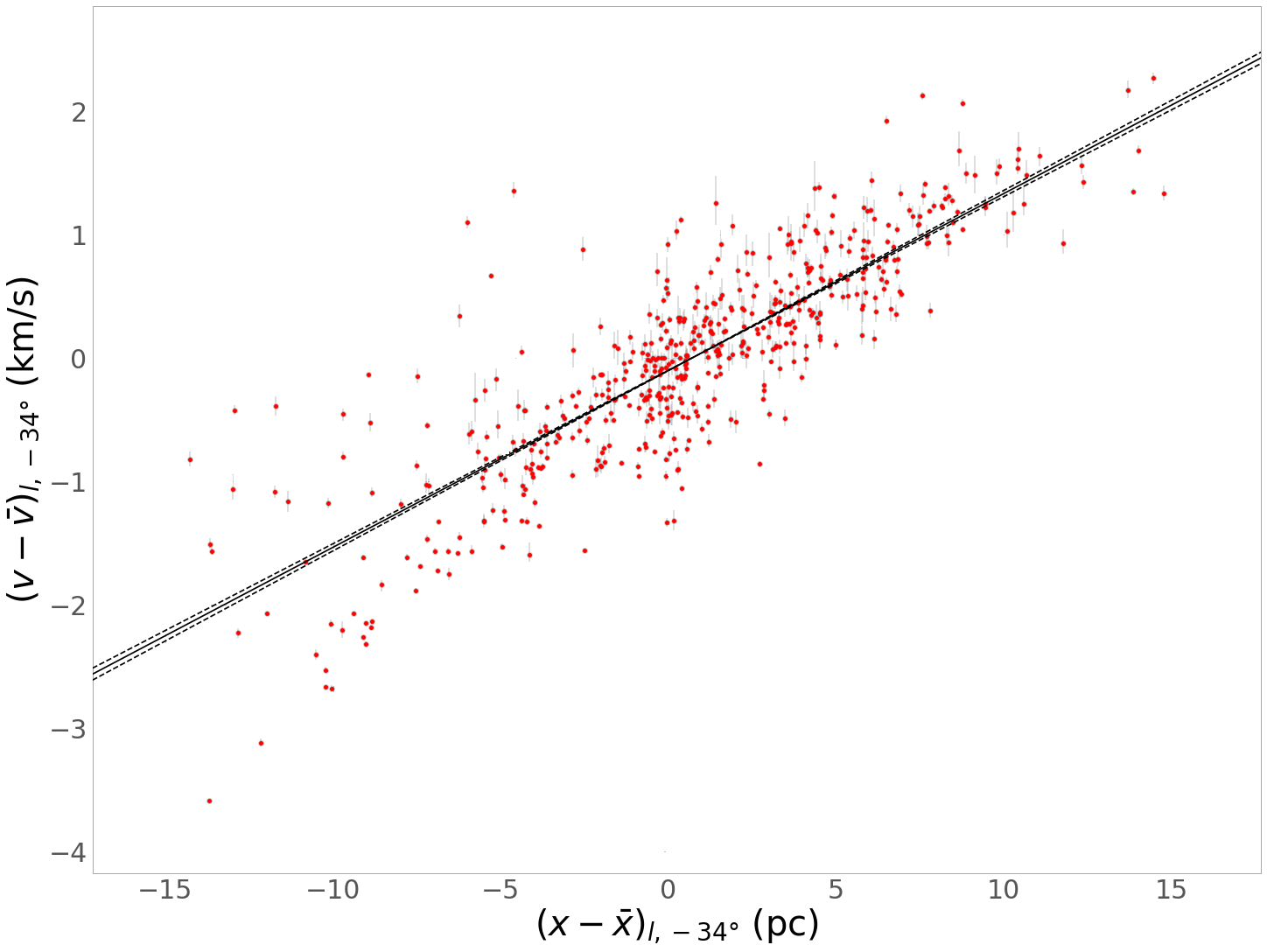} }%
    \setlength{\belowcaptionskip}{-10pt}
    \setlength{\textfloatsep}{0pt}
    \caption{Position versus expansion velocity in the direction of maximum expansion for members of $\lambda$ Ori with the best-fit linear gradient and 1$\sigma$ errors of $0.144^{+0.003}_{-0.003}$ kms$^{-1}$pc$^{-1}$, which indicates a  significant expansion trend with a corresponding expansion timescale of $6.944^{+0.148}_{-0.142}$ Myr.}%
    \label{MaxExpansion}%
\end{figure}

We find the direction of maximal expansion for $\lambda$ Ori to be oriented at -34$^\circ$ clockwise from the Galactic plane with an expansion rate of $0.144^{+0.003}_{-0.003}$ kms$^{-1}$pc$^{-1}$ (Fig.~\ref{MaxExpansion}). In Table~\ref{kinematic_table} we list the rates of expansion we obtain for $\lambda$ Ori in the direction of maximum expansion as well as the corresponding expansion timescale. 

\subsection{Expansion asymmetry}
\label{Expansion asymmetry}
In addition to identifying the maximal rate of expansion, we could also identify the minimum rate of expansion of a cluster and calculate the significance of asymmetry in the cluster's rates of expansion. The minimum rate of expansion is found to be at 80$^{\circ}$ above the Galactic plane with increasing longitude, close to the direction of Galactic latitude, with an expansion rate of $0.119^{+0.004}_{-0.004}$ kms$^{-1}$pc$^{-1}$. This is different from the maximum rate of expansion at the 5$\sigma$ significance level, providing strong evidence that expansion trends across $\lambda$ Ori are asymmetric.

Interestingly, neither the directions of maximum or minimum rate of expansion correlate exactly with the directions of maximum or minimum positional spread of cluster members (see the top panel of Fig.~\ref{ExpansionAsymmetry}) oriented at 24$^\circ$ and -64$^\circ$ clockwise to the Galactic plane respectively. This is a further indication that the cluster did not expand from a compact, isotropic volume, but likely had some degree of structural elongation in its initial configuration.

\subsection{Expansion ages}
\label{Expansion ages 2D}

As we did for the expansion rate calculated in Sect.~\ref{1Dexpansion}, we derive a kinematic age for $\lambda$ Ori from the rate of expansion in the direction of maximum expansion (Fig.~\ref{MaxExpansion}) of $6.944^{+0.148}_{-0.142}$ Myr. This is a little larger than typical isochronal age estimates from the literature \citep[$\sim4-6$ Myr;][]{zari19,cao22}, though, like the kinematic age inferred from the 1D expansion trend (Sect.~\ref{1Dexpansion}), this should be considered an upper limit, since an expansion timescale implies the time by which cluster members would trace back to a point.

\begin{figure*}
        \includegraphics[width=\textwidth]{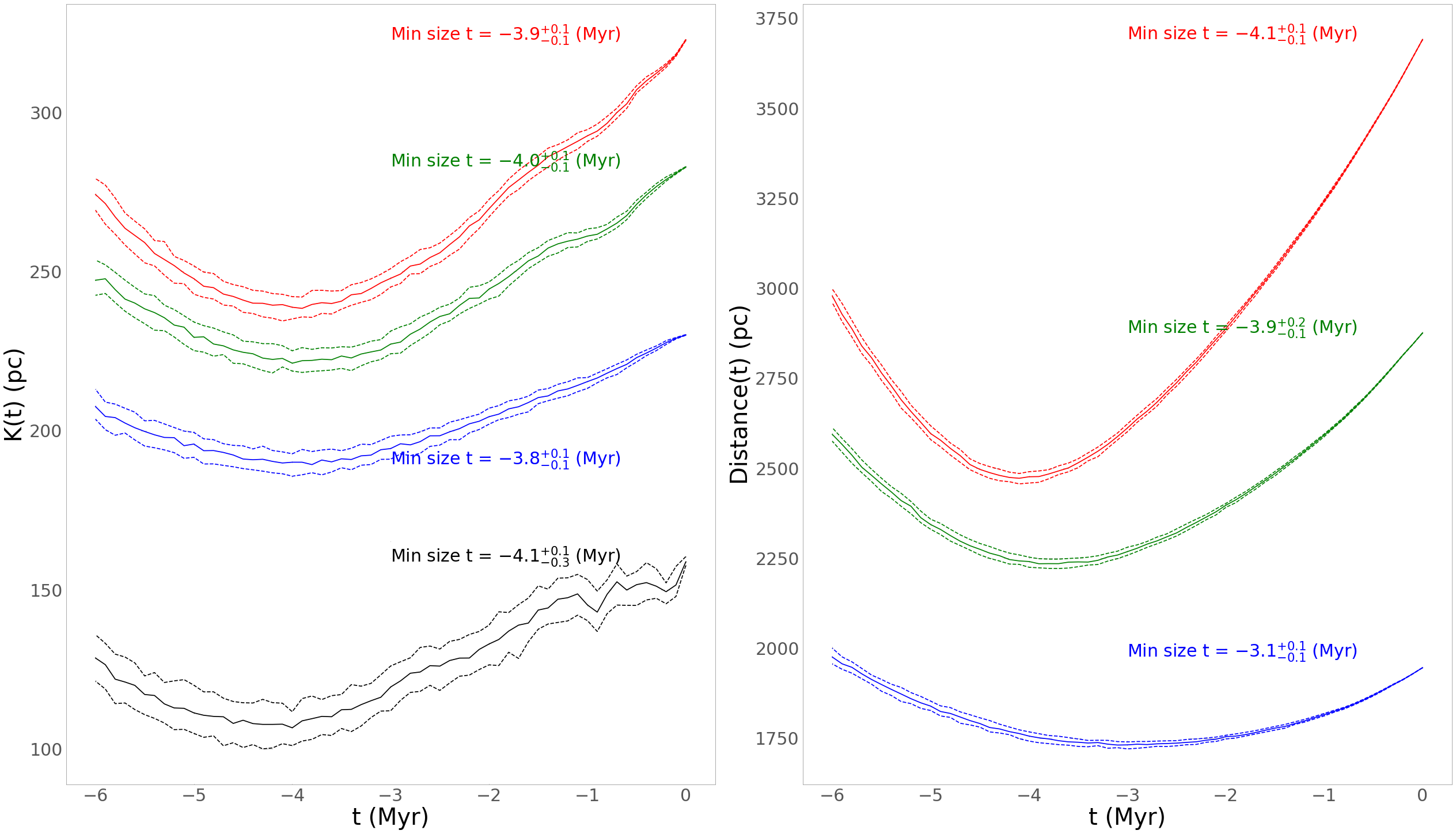}
        \setlength{\belowcaptionskip}{-10pt}
        \setlength{\textfloatsep}{0pt}
        \caption{Minimum area traceback. \textit{Left:} Minimum spanning tree total length as a function of traceback time with no filter for outliers (red), 3$\sigma$ velocity outliers removed (green), 2$\sigma$ velocity outliers removed (blue), and 32\% of the longest branches removed (black) and with their respective uncertainties. \textit{Right:} Sum of distances for each star to the association centre as a function of traceback time with no filter for outliers (red), 3$\sigma$ velocity outliers removed (green), and 2$\sigma$ velocity outliers removed (blue). }
        \label{MinAreaPlot}
\end{figure*}

\section{Rotation}
\label{rotation}

As well as the cluster expansion velocity component (Sect.~\ref{Expansion velocity}), the other component of 2D cluster velocity is that of rotation on the plane of sky, the velocity component of cluster members directed perpendicular to the cluster centre. We calculate the median cluster rotation velocity using the same approach as for the expansion velocity and find it to be $-0.054^{+0.022}_{-0.022}$ kms$^{-1}$, indicating that the $\lambda$ Ori cluster is rotating clockwise in the plane of sky as a whole.

Correlations between position and velocity in different dimensions can provide an indication of rotation in a group of stars. We repeat the same gradient fits performed in Sect. \ref{Expansion} between position and velocity in different dimensions along different position angles, similarly to the plane-of-sky expansion rates (see Sect.~\ref{Direction of Maximal Expansion}), to search for evidence of rotation anisotropy, and plot the results in Fig.~\ref{ExpansionAsymmetry} (purple). We find that direction of maximum rotation rate is oriented at -40$^\circ$ clockwise to the direction of Galactic longitude, with a rate of $0.039^{+0.009}_{-0.009}$ kms$^{-1}$pc$^{-1}$, and the direction of minimum rotation rate is oriented at 76$^\circ$ clockwise to the direction of Galactic longitude, with a rate of $-0.055^{+0.007}_{-0.007}$ kms$^{-1}$pc$^{-1}$. These rates are anisotropic at the 8$\sigma$ significance level.

\citet{guilherme23} reported a detection of significant expansion in $\lambda$ Ori (Collinder 69) but did not find a significant rotation trend. This could be because they only detect cluster average trends, regardless of direction, whereas our results indicate that the rotation rate can be either positive or negative depending on orientation, but they also do not report a measurement of average rotation rate, expansion rate or velocity, making it difficult to compare our results to theirs.

\section{Velocity dispersions}
\label{VelDispersion}

Velocity dispersions of stellar clusters are often used to investigate their dynamical state. We estimate the velocity dispersions for $\lambda$ Ori using Bayesian inference, sampling the posterior distribution with MCMC and comparing the observations to the model using a maximum likelihood \citep[see e.g.][]{Armstrong22}.

The model velocity distributions are modelled as 2-dimensional Gaussians with a total of 4 free parameters (the central velocity and velocity dispersion in each dimension). We then add an uncertainty randomly sampled from the observed uncertainty distribution in each dimension for each star.

When sampling the posterior distribution functionwe use an unbinned maximum likelihood test to compare the model and observations. The only prior we require is that velocity dispersions must be $>0$ kms$^{-1}$. This is repeated for 2000 iterations with 1000 walkers, the first half of which is discarded as burn-in. We take the median value of the posterior distribution as the best fit and the 16th and 84th percentiles as 1$\sigma$ uncertainties. Table~\ref{kinematic_table} lists the best-fit velocity dispersions and uncertainties for members of $\lambda$ Ori. 

We estimate the velocity dispersions in this way along different position angles, similarly to the plane-of-sky expansion rates (see Sect.~\ref{Direction of Maximal Expansion}), using the virtual-expansion corrected tangential velocities, and plot the results in Fig.~\ref{ExpansionAsymmetry} (green). We find that direction of maximum velocity dispersion is at 32$^{\circ}$ above the Galactic plane with increasing longitude, with a dispersion of $1.225^{+0.048}_{-0.043}$ kms$^{-1}$, and the direction of minimum velocity dispersion is oriented at -62$^\circ$ clockwise to the direction of Galactic longitude, with a dispersion of $0.771^{+0.029}_{-0.026}$ kms$^{-1}$. These dispersions are anisotropic at the 8$\sigma$ significance level. 

Anisotropic velocity dispersions are often interpreted as indicating that a young cluster is dynamically young; that is, it has not yet undergone sufficient dynamical evolution for its velocity dispersion to tend to isotropy. For OB associations where anisotropic velocity dispersions have been observed before \citep[e.g.][]{wright16,Armstrong22} it is assumed that the substructure seen in these regions also contributes to the anisotropy, which may also be the case for $\lambda$ Ori, considering the substructure identified in Sect.~\ref{AngularDispersionParameter}. 

\begin{table}
\caption{\label{kinematic_table} Kinematic properties for the $\lambda$ Ori cluster with cluster members identified by \protect\citet{cantat-gaudin20}.}
\begin{center}
{\renewcommand{\arraystretch}{1.2}
\begin{tabular}{|p{5.7cm}|p{2.0cm}|}
\hline
No. Members with RUWE $<$ 1.4 & 563 \\ 
Mean RA ($^\circ$) & $83.788^{+0.038}_{-0.037}$ \\ 
Mean Dec ($^\circ$) & $9.823^{+0.031}_{-0.031}$ \\ 
Mean $\mu_{\alpha}$ (masyr$^{-1}$) & $1.137^{+0.020}_{-0.020}$ \\ 
Mean $\mu_{\delta}$ (masyr$^{-1}$) & $-2.079^{+0.015}_{-0.014}$ \\ 
Mean l ($^\circ$) & $195.150^{+0.039}_{-0.039}$ \\ 
Mean b ($^\circ$) & $-12.045^{+0.029}_{-0.029}$ \\ 
Centre of Mass l ($^\circ$) & $195.085^{+0.012}_{-0.023}$ \\ 
Centre of Mass  b ($^\circ$) & $-11.978^{+0.006}_{-0.009}$ \\ 
Core radius (pc) & 0.44 \\
Cluster core velocity $v_{l_0}$ (kms$^{-1}$) & $4.046^{+0.079}_{-0.077}$ \\ 
Cluster core velocity $v_{b_0}$ (kms$^{-1}$) & $-0.669^{+0.045}_{-0.049}$ \\ 
2D Ellipse $\theta$ ($^\circ$) & 24.3 \\ 
2D Ellipse e & 0.695 \\ 
No. RVs & 189 \\ 
Median RV (kms$^{-1}$) & $27.47^{+0.08}_{-0.08}$ \\ 
Median LSR RV (kms$^{-1}$) & $12.37^{+0.08}_{-0.08}$ \\ 
Median Distance (pc) & $399.68^{+0.89}_{-0.89}$ \\ 
Fraction of members with $|\Delta\theta|<45^\circ$ & 0.59 \\ 
Fraction of members with $|\Delta\theta|<90^\circ$ & 0.74 \\ 
Expansion Velocity (kms$^{-1}$) & $0.711^{+0.021}_{-0.021}$ \\ 
Rotational Velocity (kms$^{-1}$) & $0.097^{+0.018}_{-0.018}$ \\ 
1D Expansion Rate (kms$^{-1}$pc$^{-1}$) & $0.181^{+0.030}_{-0.030}$ \\ 
1D Expansion Timescale (Myr) & $5.637^{+1.109}_{-0.800}$ \\ 
Max Expansion Rate (kms$^{-1}$pc$^{-1}$) & $0.144^{+0.003}_{-0.003}$ \\ 
Max Expansion Angle ($^\circ$) & $-34$ \\ 
Max Expansion Timescale (Myr) & $6.944^{+0.148}_{-0.142}$ \\ 
Min Expansion Rate (kms$^{-1}$pc$^{-1}$) & $0.119^{+0.004}_{-0.004}$ \\ 
Min Expansion Angle ($^\circ$) & $80$ \\ 
Min Expansion Timescale (Myr) & $8.403^{+0.292}_{-0.273}$ \\ 
Expansion Asymmetry ($\sigma$) & 5.0 \\ 
Max Velocity Dispersion (kms$^{-1}$) & $1.225^{+0.048}_{-0.043}$ \\ 
Max Velocity Dispersion Angle ($^\circ$) & $32$ \\ 
Min Velocity Dispersion (kms$^{-1}$) & $0.771^{+0.029}_{-0.026}$ \\ 
Min Velocity Dispersion Angle ($^\circ$) & $-62$ \\ 
Velocity Dispersion Asymmetry ($\sigma$) & 8 \\ 
Max Rotation Rate (kms$^{-1}$pc$^{-1}$) & $0.039^{+0.009}_{-0.009}$ \\ 
Max Rotation Rate Angle ($^\circ$) & $-40$ \\ 
Min Rotation Rate (kms$^{-1}$pc$^{-1}$) & $-0.055^{+0.007}_{-0.007}$ \\ 
Min Rotation Rate Angle ($^\circ$) & $76$ \\ 
Rotation Asymmetry ($\sigma$) & 8 \\ 
Minimum area traceback timescale (Myr) & $4.1^{+0.1}_{-0.1}$ \\
Median 2D Closest Approach distance for expanding members ($^\circ$) & 0.26 \\ 
Median 2D Closest Approach timescale for expanding members (Myr) & 4.73 \\ 
\hline
\end{tabular}}
\end{center}
\tablefoot{See the text for a discussion of how these quantities were derived.}
\end{table}

\section{Two-dimensional traceback}
\label{Traceback}
As we mentioned in Sects.~\ref{1Dexpansion} and \ref{Expansion ages 2D}, expansion timescales should be considered upper limit kinematic ages since they correspond to the time in the past when the cluster members would trace back to a point, given their current velocities. Alternatively, if we assume that a cluster has formed with some initial volume that is not very much smaller than its present volume, an expansion timescale will then be an overestimated kinematic age. 

Expansion timescales also implicitly assume that all the cluster members began moving away from the cluster centre at the same time, which may be a reasonable approximation in a scenario where residual gas expulsion is the dominant mechanism responsible for a cluster becoming unbound. Alternatively, we might expect that cluster members become unbound from the cluster gradually as the cluster evolves dynamically, or that the cluster forms unbound in a sparse initial configuration, and that the cluster members begin to disperse as they are formed over some non-negligible cluster-formation timescale. 

In such cases, kinematic age methods that assume a non-negligible initial volume, or which allow for stars to become unbound from the cluster at different times, might be more accurate. 

Comparing the results of expansion analysis (Sect.~\ref{1Dexpansion}) with other dynamical traceback methods can help us to distinguish between these different formation scenarios and dispersal mechanisms involved in the cluster's dynamical evolution.

\subsection{Minimum area traceback}
\label{Min Area}
We can calculate the size of a cluster at time $t$ in the past, and thus when in the past the size of the cluster would have been at its minimum. This then gives another kinematic age which assumes that the minimum size of the cluster is approximately its initial configuration.

For a sample of members of the Upper Scorpius subgroup of the Sco-Cen association, \citet{squicciarini21} performed both 2D traceback of their proper motions and 3D traceback for those with RVs, and used the total length of the minimum spanning tree between members at time $t$ to quantify the spatial coherence of the kinematic subgroups they identified. To determine the dynamical age estimates they find the time $t$ at which the total length of the minimum spanning tree is at a minimum, excluding the 10$\%$ longest branches for 2D and 32$\%$ longest branches in 3D.

\citet{quintana22} used a similar approach in their study of OB star kinematics in Cygnus, but they quantify the spatial coherence of kinematic subgroups by calculating the sum of the distance between each star and the group median position at a point in time. They then take the time in the past when the sum of distances is at a minimum as the dynamical age estimate. 

We calculate 2D Cartesian positions and velocities back in time $X(t),Y(t),U(t),V(t)$ up to 6 Myrs in the past using a linear approximation in 0.1 Myr steps. At each step we calculate the spatial coherence of the association using both the MST total length \citep{squicciarini21} and distance sum \citep{quintana22} methods. We estimate uncertainties on these using a Monte Carlo process with 1000 iterations, taking the 84th and 16th percentiles of the posterior distribution as the 1$\sigma$ uncertainties. We plot the resulting metrics as functions of traceback time $t$ in Fig.~\ref{MinAreaPlot} with different filters on the cluster members included, all members in red, 3$\sigma$ velocity outliers removed in green, 2$\sigma$ velocity outliers removed in blue, and the 32\% longest branches removed in black (Fig.~\ref{MinAreaPlot} \textit{left}). We also give the traceback times at which these metrics are minimised, with uncertainties.

In general, the resulting kinematic ages by minimum area traceback, with either MST total length or summed distances methods and with different outlier rejection criteria, are in good agreement except for the summed distances with 2$\sigma$ velocity outliers removed (blue, \textit{right}). 

The resulting past time at which the sky area of the $\lambda$ Ori cluster would have been at a minimum, as indicated by the full cluster sample (red) and 32\% longest MST branches removed (black), and thus the kinematic age of the cluster by this method, is $\tau_{\rm min\: area} = 4.1^{+0.1}_{-0.1}$ Myr. This is significantly smaller than the 1D expansion timescale of $\tau_{exp,1D} = 5.637^{+1.109}_{-0.800}$ Myr (Fig.~\ref{1D}).

According to full cluster (red) and 32\% longest MST branches removed (black) samples, the MST total length and summed distance at present ($t = 0$) are $\sim50\%$ greater than the minimum, indicating that the minimum area of the cluster would have been $\sim2.25$ times smaller than at present.

\begin{figure*}
\begin{subfigure}{0.49\textwidth}
    \includegraphics[width=\columnwidth]{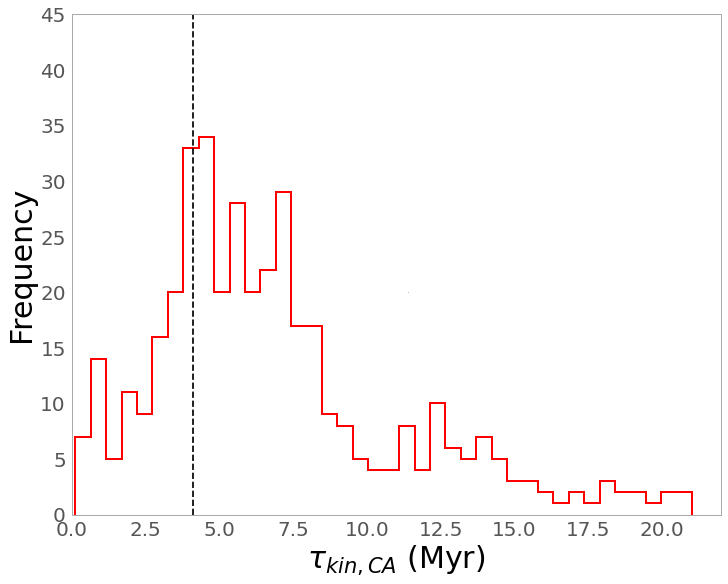}
\end{subfigure}
\begin{subfigure}{0.49\textwidth}
    \includegraphics[width=\columnwidth]{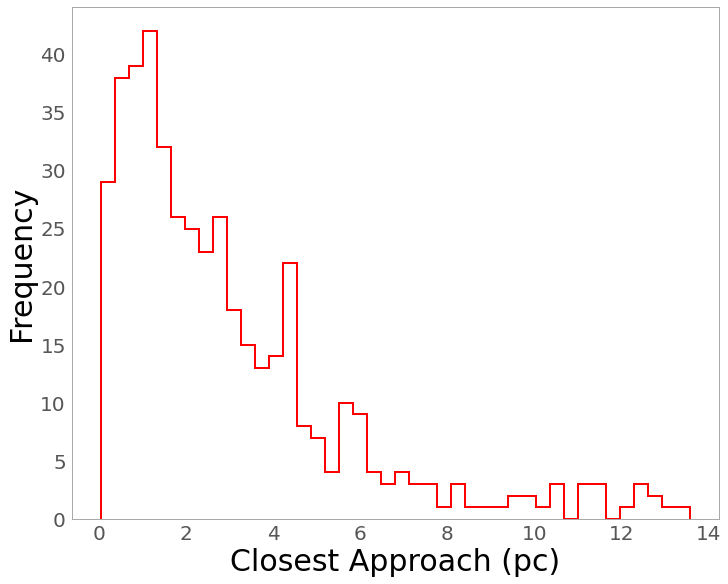}
\end{subfigure}
\caption{Traceback to closest approach. \textit{Left:} Histogram of kinematic ages for members of $\lambda$ Ori that have a positive expansion velocity. \textit{Right:} Histogram of plane-of-sky closest approach distances to the cluster centre for $\lambda$ Ori members that have a positive expansion velocity. }
\label{Histograms}
\end{figure*}

\begin{figure*} 
    {\includegraphics[width=500pt]{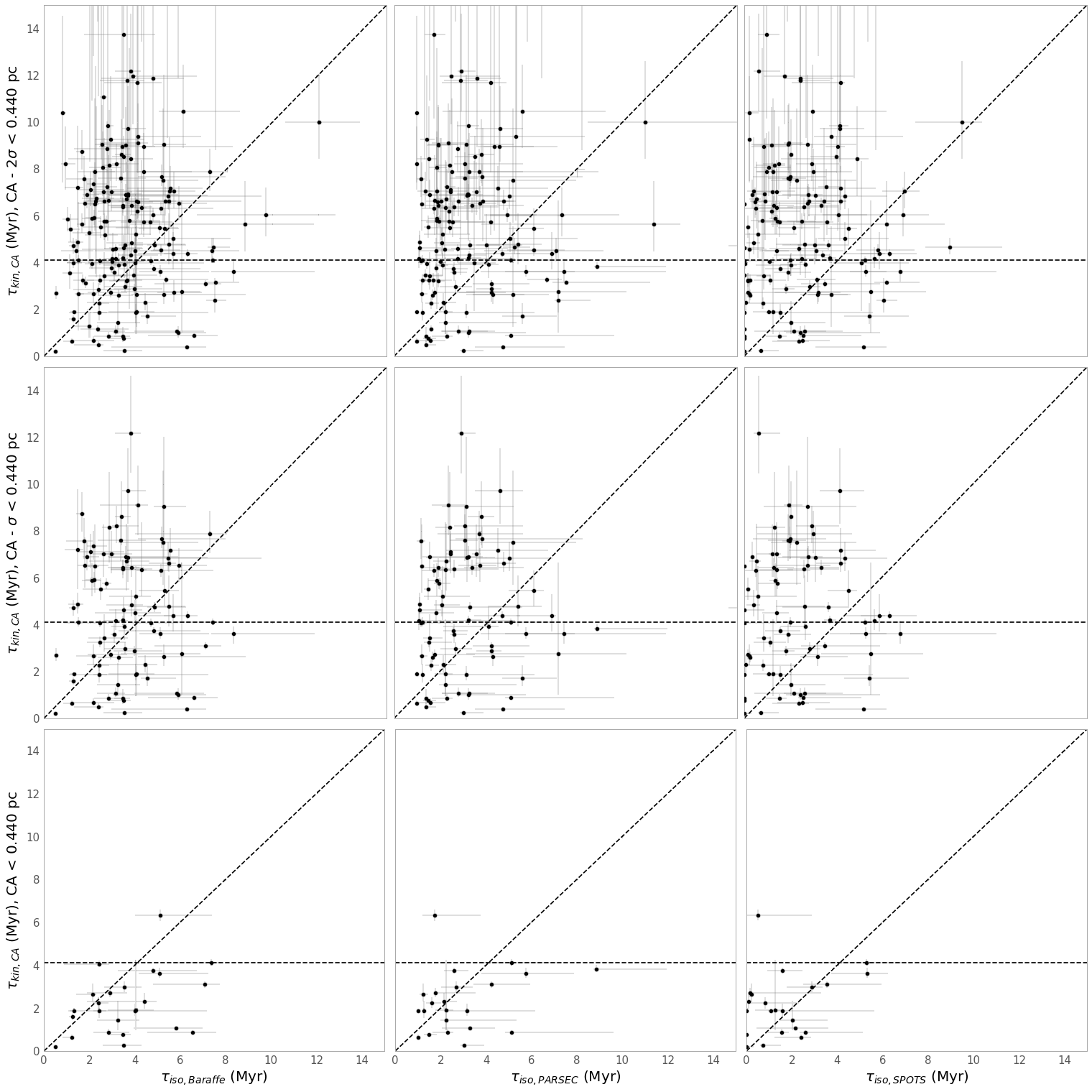} }%
    \setlength{\belowcaptionskip}{-10pt}
    \setlength{\textfloatsep}{0pt}
    \caption{Isochronal ages versus plane-of-sky traceback age for $\lambda$ Ori members whose plane-of-sky motion traces back to $< r_{c}$ pc (black) or $< 1$ pc (grey) of the centre of the cluster and whose plane-of-sky position-velocity angle ($\Delta\theta$) is within 45$^\circ$. Isochronal ages are estimated using G-RP colour (right colour), compared to \protect\citet{baraffe15,marigo17,somers20} stellar evolution models (top to bottom rows respectively) with correction for extinction and reddening.}
    \label{IsoVsTBAges}%
\end{figure*}

\subsection{Star by star}
\label{star-by-star}

For individual cluster members we can estimate a `kinematic traceback age' by  extrapolating a star's past trajectory from its present velocity to the point of closest approach to the cluster centre and calculating the time since closest approach $\tau_{kin, CA}$. This assumes that the point of closest approach to the cluster centre is the initial position of that cluster member but, unlike expansion timescales (Sect.~\ref{1Dexpansion}) and minimum area traceback (Sect.~\ref{Min Area}), does not assume that all dispersing cluster members became unbound at the same time.

In Fig.~\ref{Histograms} (\textit{left}) we show histograms of kinematic ages using the traceback to the point of closest approach $\tau_{kin, CA}$ (red), for members of $\lambda$ Ori outside the core radius $r_c = 0.44$ pc, using virtual-expansion corrected tangential velocities. We find a median kinematic age of 4.73 Myr. This is in reasonable agreement with literature ages for $\lambda$ Ori \citep[$\sim4-6$ Myr;][]{zari19,cao22}. 

In Fig.~\ref{Histograms} (\textit{right}) we also show a histogram of the closest approach distance on the sky to the cluster centre (pc) of cluster members. The median closest approach distance is found to be 0.26 $^\circ$. 

A similar approach is used by \citet{pelkonen23} to calculate ``evaporation ages'' for subgroups of the Upper Scorpius association, by selecting members of subgroups whose closest approach to the subgroup centre is within the subgroup core radius and calculating the median of the upper 30th percentile of their traceback times, after removing outliers. In this method they define a subgroup's core radius as the median distance of subgroup members from the subgroup center, effectively the half-mass radius $R_{50}$. For our sample of $\lambda$ Ori cluster members the half-mass radius $R_{50} = 6.69$ pc, much greater than the core radius of $r_c = 0.44$ pc we define, and the evaporation age calculated following the method of \citet{pelkonen23} is $\tau_{eva} = 7.87$ Myr, again much larger than either the minimum area traceback time $\tau_{\rm min\: area}$ or the expansion timescale $\tau_{exp,1D}$. However, it should be noted that this result is based on 2D traceback, in contrast to the results obtained by traceback in 3D in \citet{pelkonen23}. It remains to be seen if the evaporation age $\tau_{eva}$ of $\lambda$ Ori would change significantly with more RVs available for cluster members.

\subsection{Comparison to isochronal ages}
\label{Isochronal ages}
We also estimate ages for cluster members by interpolating between stellar evolution models using their positions on colour-absolute magnitude diagrams. We calculate \textit{Gaia} DR3 G-RP colour and absolute G magnitudes for $\lambda$ Ori cluster members and estimate extinction A$_{G}$ and reddening E(G-RP) using an approach similar to that presented in \citet{zari18}, which has previously been applied to large samples of candidate young stars in \textit{Gaia} \citep[e.g.][]{schoettler20,farias20}. We use extinction estimates from the StarHorse catalogue \citep{starhorse} for sources within the volume $81^\circ < RA < 86.5^\circ$, $7.5^\circ < Dec < 12.5^\circ$ and $2.0$ mas $< \varpi < 3.0$ mas, which contains all $\lambda$ Ori cluster members of \citet{cantat-gaudin20}. We then divide this sample into a grid of 0.25$^\circ$ x 0.25$^\circ$ x 0.2 mas cells and calculate median, 16th and 84th percentile values of A$_{G}$ and E(G-RP) per cell. We then apply these values to $\lambda$ Ori cluster members according to which cell they are located in. 

The StarHorse catalogue \citep{starhorse} provides A$_{G}$ estimates for 225 $\lambda$ Ori cluster members of \citet{cantat-gaudin20}, which range from -0.03 to 1.51 with a median of 0.30, whereas our interpolated values from a grid of the StarHorse catalogue range from 0.20 to 0.50 with a median of 0.32. 

We also note that 53 candidate $\lambda$ Ori cluster members of \citet{cao22} match with StarHorse sources within the aforementioned volume. \citet{cao22} estimated extinctions $A_{V}$ for candidate $\lambda$ Ori cluster members with spectral types and $T_{eff}$ estimates derived from APOGEE spectra. For these 53 sources we perform least-squares fitting to find the gradient of a linear best-fit between extinctions $A_{V}$ from \citet{cao22} and \citep{starhorse}. We use a Monte Carlo approach with random sampling of $A_{V}$ uncertainties, using the 16th and 84th percentile values of $A_{V}$ from \citep{starhorse} as 1$\sigma$ uncertainties on the median, for each source, which we repeat for 10 000 iterations. We find a median linear gradient of $1.079^{+0.212}_{-0.203}$, with a median scatter of the data around the best-fit line of $0.906^{+0.075}_{-0.084}$ mag. We calibrate the extinction values in our grid using this median linear gradient, propagating the uncertainties via a Monte Carlo approach. 

Using these values, we infer ages for cluster members by linear interpolation between sets of \citet{baraffe15}, \citet{marigo17} and \citet{somers20} isochrones, respectively. We repeat this interpolation for 1000 iterations per cluster member, each time with randomly sampled uncertainties for LOS distance and A$_{G}$. The random factor of uncertainty applied to A$_{G}$ is then also applied to E(G-RP) uncertainty each iteration, to maintain a consistent extinction law. The 50th percentile value of the posterior distribution of the interpolated ages is taken as the median, and the 16th and 84th percentile values are then taken as the uncertainties. 

Stellar evolution models vary both in the physics included (e.g. magnetic or non-magnetic models) and in the initial conditions assumed. Thus, the inferred $\tau_{iso}$ varies significantly between models on a star-by-star basis and depending on their location in the G-RP versus M$_{G}$ colour-magnitude diagram.  

In Fig.~\ref{IsoVsTBAges} we plot $\tau_{iso}$ against $\tau_{kin,CA}$ for unbound cluster members whose plane-of-sky motion traces back to within the core radius ($r_{c} = 0.44$ pc) of the cluster within 2$\sigma$ closest approach uncertainty, within 1$\sigma$ uncertainty and where the median closest approach distance is within the core radius, respectively in descending rows. There is significant spread in $\tau_{iso}$ with larger uncertainties for sources with greater $\tau_{iso}$, reflecting the uncertainties in the stellar evolution models at this stage of PMS evolution. 
 
A Pearson correlation test performed on $\tau_{iso}$ and $\tau_{kin,CA}$ for sources with median closest approach distance $< r_{c}$ yields correlation coefficients of $0.417$ with $p$ value of $0.048$, $0.342$ with $p$ value of $0.110$ and $0.308$ with $p$ value of $0.153$, using \citet{baraffe15}, \citet{marigo17} and \citet{somers20} isochrones respectively, indicating a low to moderate correlation. This correlation becomes even less significant when including sources with closest approach distances that only reach the core radius within 1$\sigma$ or 2$\sigma$ uncertainties. 

We also calculated the median $\tau_{iso}$ for upper and lower quartiles of $\tau_{kin,CA}$ for cluster members in each panel of Fig.~\ref{IsoVsTBAges}. For cluster members whose median closest approach distance $< r_{c}$ (Fig.~\ref{IsoVsTBAges}~\textit{bottom row}), the mean $\tau_{iso}$ for the upper and lower quartiles are $3.739^{+0.993}_{-1.470}$ and $3.025^{+1.055}_{-1.498}$ respectively, using \citet{baraffe15} models, $2.419^{+0.781}_{-0.801}$ and $2.761^{+0.839}_{-0.916}$ using \citet{marigo17} models and $1.649^{+0.971}_{-0.957}$ and $1.972^{+0.888}_{-0.964}$ using \citet{somers20} models. Similarly, there is no significant difference in the mean $\tau_{iso}$ for the upper and lower quartiles of $\tau_{kin,CA}$ of cluster members in Fig.~\ref{IsoVsTBAges} in the top and middle rows.

This lack of clear correlation between $\tau_{iso}$ and $\tau_{kin,CA}$ is due in part to large uncertainties in $\tau_{iso}$ estimates, which in turn are mostly due to uncertainties in estimates of A$_{G}$ and E(G-RP). However, apart from the uncertainties in estimating $\tau_{iso}$ and $\tau_{kin,CA}$, there are physical reasons for a scatter around unity. Cluster members may not become unbound from the cluster until several Myr into their evolution; thus we would expect to see $\tau_{iso} > \tau_{kin,CA}$. On the other hand, despite our efforts to filter the sample there may still be sources whose origin is outside the cluster core, so a kinematic age based on closest approach to the cluster centre $\tau_{kin,CA}$ would overestimate age, which is likely the case for cluster members with $\tau_{iso} < \tau_{kin,CA}$ in Fig.~\ref{IsoVsTBAges} top and middle rows.

\begin{figure} 
    \subfloat{{\includegraphics[width=245pt]{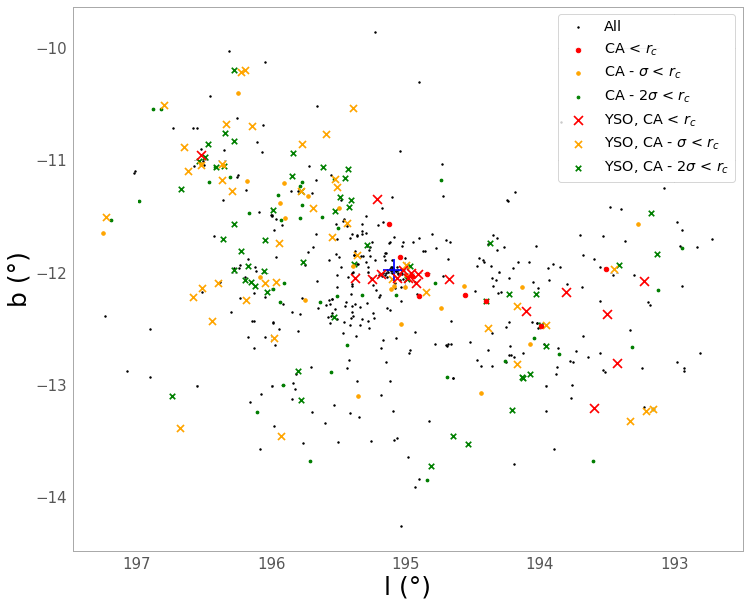} }}%
    \qquad
    \subfloat{{\includegraphics[width=245pt]{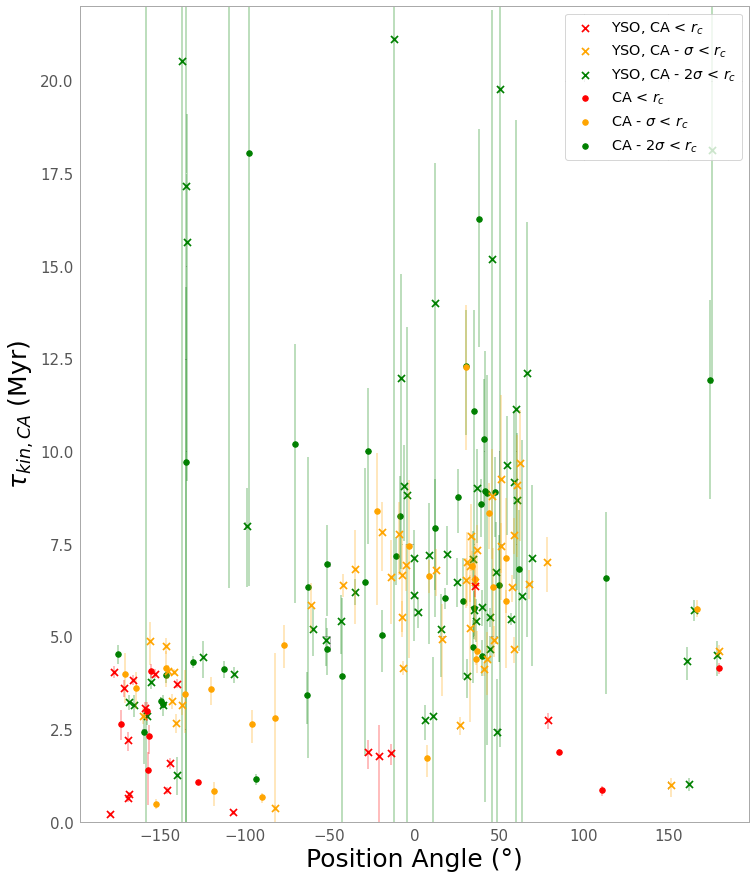} }}%
    \setlength{\belowcaptionskip}{-10pt}
    \setlength{\textfloatsep}{0pt}
    \caption{Properties of candidate ejected stars. \textit{Top:} Galactic coordinates of the $\lambda$ Ori cluster members (black) with members that traceback to within the core radius ($CA < r_c$) in red, members that traceback to the core radius within 1$\sigma$ ($CA - \sigma < r_c$) in yellow, members that traceback to the core radius within 2$\sigma$ ($CA - 2\sigma < r_c$) in green, and the position of the cluster centre denoted with a blue cross. Cluster members consistent with traceback to the core radius that match with the \textit{Gaia} DR3 varYSO catalogue \protect\citep{marton23} are plotted with X's.  \textit{Bottom:} Position angle ($^{\circ}$) around the cluster centre from the direction of increasing Galactic longitude for $\lambda$ Ori cluster members that traceback to within the core radius ($CA < r_c$) in red, members that traceback to the core radius within 1$\sigma$ ($CA - \sigma < r_c$) in yellow, and members that traceback to the core radius within 2$\sigma$ ($CA - 2\sigma < r_c$) in green against a timescale of 2D traceback to the point of closest approach to the cluster centre $\tau_{kin,CA}$. Cluster members consistent with traceback to the core radius that match with the \textit{Gaia} DR3 varYSO catalogue \protect\citep{marton23} are plotted with X's.}
    \label{Ejectees}%
\end{figure}

\subsection{Oldest cluster core escapees}
\label{core escapees}

The largest kinematic traceback ages $\tau_{kin,CA}$ of unbound cluster members, consistent with originating from the core of the cluster, can also give a lower-limit age for the cluster. This approach has been previously applied to confirmed samples of dynamically ejected runaway stars from young clusters \citep[e.g.][]{stoop23,fajrin24}.

In the lower row of Fig.~\ref{IsoVsTBAges} we note that $\tau_{kin,CA}$ for these cluster members are all $< \tau_{\rm min\: area}$, with one exception. \textit{Gaia} DR3 ID 3336145675718723968 has $\tau_{kin,CA} = 6.34^{+0.31}_{-0.26} Myr$, a 2D closest approach distance of $0.34^{+0.34}_{-0.24}$ pc, is flagged as \textit{Gaia} DR3 VarYSO \citep{marton23} and has an RV from the APOGEE survey, calibrated in the \textit{Survey of Surveys} \citep{tsantaki22}, of $29.56 \pm 0.27$ kms$^{-1}$. With this RV we can calculate it's trajectory in 3D using the approach employed by \citet{fajrin24} for candidate runaway stars, in order to verify the 2D traceback results. We transform the observed astrometry of this source into Galactic Cartesian coordinates $X, Y, Z$ and velocities $U, V, W$, and do the same for the reference frame of the cluster, using the mean cluster coordinates, proper motions, distance and RV as given in Table~\ref{kinematic_table}. Then we calculate the 3D closest approach distance $D_{min,3D}$ and 3D traceback time $\tau_{min,3D}$ using Eqs. 1 and 2 of \citet{fajrin24}. We find $D_{\rm min,3D} = 4.83 \pm 2.77$~pc and $\tau_{min,3D} = 3.23 \pm 0.58$~Myr, indicating that this source likely did not originate from the core of $\lambda$ Ori, in contrast with the 2D traceback. However, this $\tau_{\rm min,3D}$ is in better agreement with the source's isochronal ages $\tau_{\rm iso}$ than the 2D traceback time $\tau_{\rm kin,CA}$.

Then, excluding \textit{Gaia} DR3 ID 3336145675718723968, the largest $\tau_{\rm kin,CA}$ values for for unbound cluster members whose plane-of-sky motion traces back to within the core radius ($r_{c} = 0.44$ pc) are $4.1 \pm 0.1$~Myr, in close agreement with the minimum area traceback age $\tau_{\rm min\: area}$ and isochronal ages for the cluster from the literature \citep{zari19,cao22}. 

In Fig.~\ref{Ejectees}~\textit{Top} we plot the positions of cluster members which traceback to within the core radius ($CA < r_c$) in red, members that traceback to the core radius within 1$\sigma$ ($CA - \sigma < r_c$) in yellow, members that traceback to the core radius within 2$\sigma$ ($CA - 2\sigma < r_c$) in green, and the position of the cluster centre denoted with a blue cross. Cluster members consistent with traceback to the core radius that match with the \textit{Gaia} DR3 varYSO catalogue \citep{marton23} are plotted with 'x's, and we find that 18/27 red points, 44/68 yellow points and 56/97 green points are varYSOs. 

In Fig.~\ref{Ejectees}~\textit{Bottom} we plot the position angles and traceback ages $\tau_{kin,CA}$ for these candidate ejected cluster members. It is apparent that the majority of these cluster members are preferentially located at position angles $\sim30-40^{\circ}$ and $\sim-150^{\circ}$, which are in good agreement with the directions of maximum spatial spread and maximum velocity dispersion for the whole cluster (see Fig.~\ref{ExpansionAsymmetry}), though the red points in particular are found preferentially on one side of the cluster centre. However, the fact that the majority of these candidate ejectees match with the \textit{Gaia} DR3 varYSO catalogue indicate that this stream of stars located at a particular position angle are not simply field star contaminants following Galactic differential rotation. 

\begin{figure} 
    {\includegraphics[width=\columnwidth]{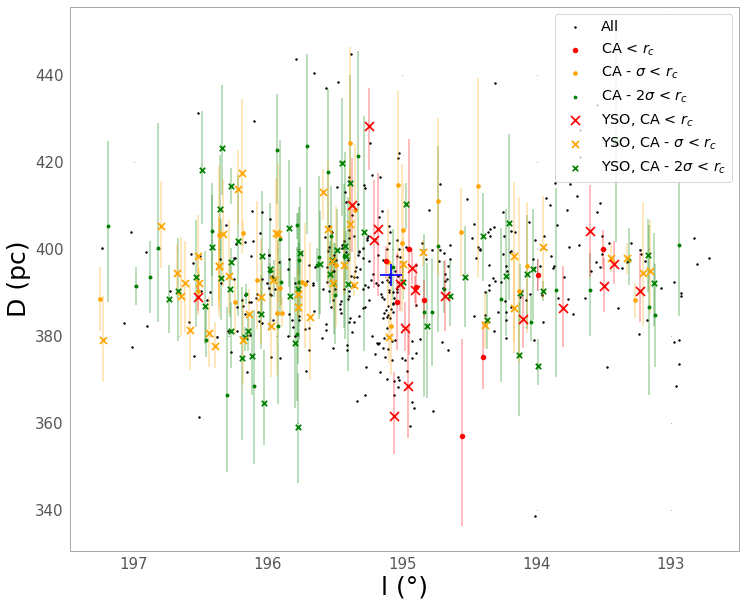} }%
    \setlength{\belowcaptionskip}{-10pt}
    \setlength{\textfloatsep}{0pt}
    \caption{Galactic longitude versus distance \protect\citep{bailerjones21} of $\lambda$ Ori cluster members (black) with members that traceback to within the core radius ($CA < r_c$) in red, members that traceback to the core radius within 1$\sigma$ ($CA - \sigma < r_c$) in yellow, members that traceback to the core radius within 2$\sigma$ ($CA - 2\sigma < r_c$) in green, and the position of the cluster centre denoted with a blue cross. Cluster members consistent with traceback to the core radius that match with the \textit{Gaia} DR3 varYSO catalogue \protect\citep{marton23} are plotted with X's. }%
    \label{Collinder_69_ejectees_l_D}%
\end{figure}

If we consider the line-of-sight distances of the candidate ejected cluster members we find that the candidates which trace back closest to the cluster core (red) occupy a large distance range, from $356.93$ to $428.37$ pc (see Fig.~\ref{Collinder_69_ejectees_l_D}). So although they appear close the cluster core on the plane of sky and have small $\tau_{\rm kin,CA}$, their 3D distances are much larger, and thus the $\tau_{\rm kin,CA}$ are likely underestimates of their kinematic ages, though RVs for these sources would be necessary to confirm their status as ejectees in 3D. 

Cluster members dynamically ejected by either the binary supernova scenario or a strong dynamical encounter of a multiple system are not expected to be ejected in a preferred direction. Thus, we would expect their distribution of position angles to be random. Recently, \citet{polak24} presented results from simulations of the formation of massive star clusters using TORCH, and demonstrated a scenario whereby the merging of sub-clusters of newly formed stars could produce groups of ejected 'runaways' with highly anisotropic distributions, which they dub the 'sub-cluster ejection scenario' (SCES). Stars ejected in such a scenario are expected to have similar ejection timescales and velocities, so this scenario could be invoked to explain the group of candidate ejected cluster members at position angle $\sim-150^{\circ}$ and traceback ages $\tau_{kin,CA} \sim 3.5 - 4$ Myr (Fig.~\ref{Ejectees}~\textit{Bottom}). A single sub-cluster merger event cannot account for the whole range of traceback ages among sources plotted in Fig.~\ref{Ejectees}, but ejections via other scenarios, such as strong dynamical encounters, can then account for the remainder of the position angle - traceback age distribution. 

\subsection{Age spread}
\label{age spread}

We look for evidence of significant spread in each of $\tau_{kin,CA}$ and $\tau_{iso}$, for the cluster members plotted in Fig.~\ref{IsoVsTBAges}. We calculate the difference between $\tau$ and the sample mean $\tau$ and normalize by the uncertainty of $\tau$ for each cluster member, after removing the largest 10\% as outliers in each case, to calculate the spread of ages $\sigma_{\tau_{kin,CA}}$ and $\sigma_{\tau_{iso}}$.

For cluster members whose closest approach distance comes within 2$\sigma$ of the core radius (Fig.~\ref{IsoVsTBAges}~\textit{top row}), the mean $\sigma_{\tau_{kin,CA}}$ is 3.3, and for $\sigma_{\tau_{iso}}$ the mean is 2.1, 2.8 and 2.1 for \citet{baraffe15}, \citet{marigo17} and \citet{somers20} models respectively. 

For cluster members whose closest approach distance comes within 1$\sigma$ of the core radius (Fig.~\ref{IsoVsTBAges}~\textit{middle row}), the mean $\sigma_{\tau_{\rm kin,CA}}$ is 4.1, and for $\sigma_{\tau_{iso}}$ the mean is 1.7, 1.7 and 1.3 for \citet{baraffe15}, \citet{marigo17} and \citet{somers20} models respectively. 

For cluster members whose median closest approach distance comes within the core radius (Fig.~\ref{IsoVsTBAges}~\textit{bottom row}), the mean $\sigma_{\tau_{\rm kin,CA}}$ is 6.4, and for $\sigma_{\tau_{iso}}$ the median is 0.9, 0.9 and 0.8 for \citet{baraffe15}, \citet{marigo17} and \citet{somers20} models respectively. 

This indicates that the spread in $\tau_{\rm kin,CA}$ is significant above the uncertainties, particularly for the cluster members with small closest approach distances, consistent with originating within the cluster core radius (of 6$\sigma$ significance). These also tend to have smaller uncertainties on $\tau_{kin,CA}$, whereas spread in $\tau_{\rm iso}$ is less significant for all models, as the uncertainties in $\tau_{\rm iso}$ remain relatively large for all but the youngest ($\sim$1 Myr) members. If these candidate ejectees can be confirmed to be consistent with originating from the cluster core in 3D with RV measurements, the significant spread in $\tau_{\rm kin,CA}$ would indicate a series of dynamical ejections from the cluster core over the $\sim4$ Myr lifetime of the cluster.

\section{Discussion}

The structural analysis of the $\lambda$ Ori cluster (Sect.~\ref{Qparametersection}) yielded a $Q$-parameter value of 0.806 (see Fig.~\ref{Qparameter}), which is between the values typically considered to indicate either substructure ($Q<0.7$) or a smooth, centrally concentrated distribution ($Q>0.9$). The ADP analysis (Sect.~\ref{AngularDispersionParameter}) revealed that the central core of the $\lambda$ Ori cluster is smoothly distributed ($\delta_{\rm ADP,e,N}(50)\sim1$), while significant substructure remains in the cluster outskirts ($\delta_{\rm ADP,e,N}(r)\sim2-3$). For a system such as the $\lambda$ Ori cluster, where different regions exhibit significant differences in the distributions of members, the ADP, which measures structure as a function of radius, reveals much more detail than the $Q$-parameter, which gives only an average measure of structure across the entire population.

Figure 8 of \citet{wu17} show values of $\delta_{\rm ADP,e,N}(r)$ for simulations of clusters formed by GMC collisions with a variety of initial parameters. The ADP values we find for the $\lambda$ Ori cluster (Fig.~\ref{ADP}) vary significantly as a function of radial distance, but remain within the range $0.9 < \delta_{\rm ADP,e,N}(r) < 2.9$, which is in best agreement with the B-1-1M-nocol model of \citet{wu17} after 5 Myr, though the scales of $R$ (pc) are very different. In their other models, such as d-1-col and B-1-col, $\delta_{\rm ADP,e,N}(r)$ is often $>3$, but the average values are also comparable to ours. For model B-1-1m-col, however, the values of $\delta_{\rm ADP,e,N}(r)$ are always $>3$, making this the least compatible model with our results. It is also worth noting that, in most models of \citet{wu17}, $\delta_{\rm ADP,e,N}(r)$ is close to a minimum in the innermost sectors, which also matches our results for $\lambda$ Ori (see Fig.~\ref{ADP}).

Some of the substructure outside the cluster centre identified in Sect.~\ref{AngularDispersionParameter} likely correlates to the subgroups identified in previous studies, often labelled as B30 and B35 \citep{dolan01,kounkel18,cao22}. \citet{cao22} found evidence to suggest that these subgroups are younger than the central cluster core of $\lambda$ Ori. However, the membership determination of \citet{cantat-gaudin20} does not distinguish between these subgroups, but finds that their astrometric properties are sufficiently similar to be classified as one cluster. We refrain from attempting to separate these subgroups in our analysis because we wish to analyse substructure across the region as a whole (Sect.~\ref{structure}), because determination of the boundaries between and membership of subgroups can be quite arbitrary, being highly sensitive to the methods used (HDBSCAN, GMM, etc.) and because we intend to apply these structural and kinematic analyses to a larger sample of nearby young clusters, where comparison of results across the sample will be more effective if cluster membership has been determined in a homogeneous way for all clusters. Thus, we selected our cluster membership sample only on the basis of the \citet{cantat-gaudin20} catalogue.

$\lambda$ Ori shows significant evidence of expansion as indicated by multiple metrics. We find a significantly positive median expansion velocity of $v_{out} = 0.711^{+0.021}_{-0.021}$ kms$^{-1}$, which is a reasonable value for a young, dispersing cluster when comparing to results from \citet{kuhn19}, for example. We also calculate velocity-position angles $\Delta\theta$ for cluster members and find that 74\% have $|\Delta\theta| < 90^{\circ}$ and so are consistent with moving away from the cluster center, and 59\% have $|\Delta\theta| < 45^{\circ}$. We find that the distribution of $\Delta\theta$ has a significant peak around $0^{\circ}$ and we also find the proportion of cluster members consistent with moving outwards from the cluster centre increases with distance from the centre $R$ (see Fig.~\ref{Vangle}). We calculate significantly positive rates of expansion, both in 1D (Fig.~\ref{1D}) and in different directions across the cluster (Fig.~\ref{ExpansionAsymmetry}). 

In particular, we find that the plane-of-sky expansion rates are significantly anisotropic, with the maximum rate of expansion of $0.144^{+0.003}_{-0.003}$ kms$^{-1}$pc$^{-1}$ directed at $-34^{\circ}$ below the Galactic plane in the direction of increasing longitude. This maximum rate of expansion then infers an expansion timescale of $6.944^{+0.148}_{-0.142}$, tracing the cluster members back to their most compact 1D configuration in that dimension. Notably, this is significantly greater than the minimum area traceback timescale $\tau_{\rm min\:area} = 4.1^{+0.1}_{-0.1}$ Myr (Sect.~\ref{Min Area}), which would indicate that the $\lambda$ Ori cluster began expanding from an initially large and sparse configuration. This is further supported by considering that the MST total length and summed distances of the minimum area traceback at present ($t=0$) are only $\sim$50\% greater than at the minimum configuration (Fig.~\ref{MinAreaPlot}). This would also explain why many cluster members have large closest approach distances and $\tau_{\rm kin,CA}$ (Fig.~\ref{Histograms}).

The close agreement between the minimum area traceback timescale $\tau_{\rm min\: area} = 4.1^{+0.1}_{-0.1}$ Myr, maximum $\tau_{kin,CA}$ for candidate ejectees from the cluster core (Sect.~\ref{Ejectees}) and isochronal ages, either from literature \citep{kounkel18,zari19,cao22} or our median $\tau_{iso}$ estimates for individual cluster members (Sect.~\ref{Isochronal ages}) suggest that the $\lambda$ Ori cluster began expanding very soon after the beginning of star formation. This is in contrast to the significant difference between kinematic ages and isochronal ages found by \citet{miret-roig24} for subgroups of Upper Scorpius and other nearby stellar associations, which they interpret as an 'embedded phase' in the early evolution of these groups, where newly formed stars remained bound within their parent molecular clouds for $\sim$5 Myr before beginning to expand.

Overall, these findings suggest that $\lambda$ Ori as a whole did not simply form as a dense monolithic cluster which began expanding after the dispersal of its parent molecular cloud, but instead formed in a sparse, substructured configuration in a large volume, similar to what has been found in nearby OB associations \citep[e.g. Vela OB2; ][]{Armstrong22}. It may, therefore, be more appropriate to consider $\lambda$ Ori as a low-mass association, rather than a classical ``cluster''.

Considering the spatial substructure, the asymmetry of the expansion trends and the large spread in $\tau_{kin,CA}$, the proposed scenario of expansion in $\lambda$ Ori being triggered by a single supernova event \citep{kounkel18} seems unlikely. Rather, this evidence would better support a scenario where stars form in a molecular cloud with substructure following filaments or an expanding shell \citep[e.g, ][]{inutsuka15} and inherit their initial kinematics from the gas, but become unbound once the gas is dispersed.

\citet{wright23} measured expansion trends and velocity dispersions for a sample of 111 spectroscopically confirmed YSO $\lambda$ Ori cluster members with RVs from the \textit{Gaia}-ESO survey. They also found evidence for asymmetric expansion of 4$\sigma$ significance in their analysis, but with maximum and minimum rates of $0.20^{+0.03}_{-0.04}$ kms$^{-1}$pc$^{-1}$ and $-0.01^{+0.03}_{-0.02}$ kms$^{-1}$pc$^{-1}$, as opposed to our maximum and minimum rates of $0.144^{+0.003}_{-0.003}$ kms$^{-1}$pc$^{-1}$ and $0.119^{+0.004}_{-0.004}$ kms$^{-1}$pc$^{-1}$. Notably, the direction of maximum expansion they found at $75^{\circ}$ is close to the direction of minimum expansion we find.

\citet{wright23} measured an expansion velocity $v_{out} = 0.24^{+0.06}_{-0.01}$ kms$^{-1}$, which is significantly lower than our measurement of $v_{\rm out} = 0.711^{+0.021}_{-0.021}$ kms$^{-1}$.

The velocity dispersions we measure are also significantly greater than those measured by \citet{wright23} for members of $\lambda$ Ori observed as part of the \textit{Gaia}-ESO survey, of $\sigma_{\mu_{\alpha}}$ = $0.52^{+0.06}_{-0.04}$ kms$^{-1}$ and $\sigma_{\mu_{\delta}}$ = $0.33^{+0.04}_{-0.03}$ kms$^{-1}$, whereas our minimum velocity dispersion is $\sigma_{min}$ = $0.771^{+0.029}_{-0.026}$ kms$^{-1}$.

The differences between these results are likely due to the difference in the samples of cluster members used, 111 \textit{Gaia}-ESO targets compared to 563 in the sample we use from \citet{cantat-gaudin20}. The greater number of cluster members in our sample is also the likely cause of the greater precision in our expansion rates and velocity dispersions. The \textit{Gaia}-ESO sample also separates the $\lambda$ Ori cluster from B30 and B35 subgroups, which the clustering of \citet{cantat-gaudin20} does not distinguish. 

We plan to apply these kinematic analysis techniques to a larger sample of well-populated nearby young clusters whose membership has been determined in the same way \citep{cantat-gaudin20}. We will then compare the kinematic trends across the sample and look for correlations with cluster age, mass, etc.

With precise RVs available for more cluster members it would be possible to extend much of this analysis to 3D; to measure the expansion (an)isotropy and the direction of maximum expansion in 3D. It would also be possible to apply 3D traceback methods, such as those presented by \citet{miret-roig20,squicciarini21,Armstrong22,kerr22,quintana22} or \citet{crundall19}, that can be used to determine kinematic ages for expanding groups of stars by estimating when in the past they would have been at their minimum volume, which is assumed to be their approximate initial configuration.

\section{Summary}
Here we summarise our results.
\begin{itemize}
    \item We cross matched high-fidelity members of the nearby young cluster $\lambda$ Ori based on \textit{Gaia} astrometry \citep{cantat-gaudin20} with the catalogue of calibrated RVs from \citet{tsantaki22} in order to obtain a sample of 189 sources with six-parameter astrometry and 374 with five-parameter astrometry.
    \item We analysed the plane-of-sky spatial structure of the cluster using the $Q$-parameter \citep{cartwright04} and ADP \citep{dario14} and found that while there is a smooth, likely bound core at the centre of the cluster, there still exists significant substructure away from the centre among the unbound cluster members, indicating that the sparse halo of cluster members outside the dense core are still dynamically young. 
    \item We investigated the radial density profile of the cluster and found a core radius of $r_{c} = 0.44$ pc, defined as the radius at which the density is half the peak density. This core radius is relatively small compared to the radial distances $R$ (pc) of the cluster halo, which in this cluster membership sample extend up to $\sim16$ pc away from the cluster centre. While the stellar density of the cluster core is a factor of roughly ten greater than in the halo, only 36 members are located within it.
    \item We transformed the proper motions of the cluster members into tangential velocities using available RVs to correct for the `virtual expansion' effect. We looked for evidence of expansion using several methods:
    \begin{itemize}
        \item We measured the mean `expansion velocity', that is, the mean velocity component directed outwards from the cluster centre $v_{\rm out}$. We found $v_{\rm out} = 0.710^{+0.022}_{-0.022}$ kms$^{-1}$, which provides significant evidence that the cluster is expanding as a whole.
        \item We calculated the fraction of cluster members with relative velocities directed outwards from the cluster centre. We found that 74\% of the members have $|\Delta\theta| < 45^\circ$ and that 59\% of the members have $|\Delta\theta| < 90^\circ$, with the peak of the distribution of $\Delta\theta$ consistent with being at $0^\circ$, indicating that a majority of the cluster members are moving radially outwards from the centre. 
        \item We measured the 1D rate of expansion by fitting a linear gradient to the radial distance from the centre $R$ (pc) against $v_{\rm out}$ (kms$^{-1}$). We found a best-fitting linear gradient of $0.181^{+0.030}_{-0.030}$ kms$^{-1}$pc$^{-1}$, which is evidence for cluster-wide expansion at the 6$\sigma$ significance level. This rate of expansion also yielded an expansion timescale of $5.637^{+1.109}_{-0.800}$ Myr, which is in agreement with literature age estimates \citep{kounkel18,zari19,cao22}.
    \end{itemize}
    \item We found that the expansion rates measured in different orientations across the plane of sky, though always significantly positive, are significantly asymmetric. We found that the difference between the maximum and minimum rates of expansion is of 5$\sigma$ significance.
    \item Since the expansion is asymmetric, there must be a certain direction in which the rate of expansion is at a maximum. We found that the direction of maximum expansion is at $-34^{\circ}$ below the Galactic plane, with a rate of $0.144^{+0.003}_{-0.003}$ kms$^{-1}$pc$^{-1}$, while the direction of minimum expansion is at $80^{\circ}$ above the Galactic plane, with a rate of $0.119^{+0.004}_{-0.004}$ kms$^{-1}$pc$^{-1}$. 
    This differs with the direction of maximum expansion found by \citet{wright23} of $75^{\circ}$, which is closer to our direction of minimum expansion. This is likely due to the large difference in cluster membership samples used to analyse kinematic trends and highlights the need for more complete cluster membership for accurate analysis of kinematic trends.
    \item We also found that the velocity dispersion on the plane of sky is significantly asymmetric, with an 8$\sigma$ significance between the maximum and minimum and with the direction of maximum velocity dispersion being well correlated with the direction of the greatest spatial spread.
    \item By inverting the maximum rate of expansion, we calculated an expansion timescale of $6.944^{+0.148}_{-0.142}$ Myr, which is significantly greater than the timescale calculated from the 1D expansion gradient. 
    \item We also looked for evidence of cluster rotation in the plane of sky by fitting linear gradients to position-perpendicular velocity trends. We found a maximum rotation rate of $0.039^{+0.009}_{-0.009}$ kms$^{-1}$pc$^{-1}$ directed at $-40^{\circ}$ below the Galactic plane and a minimum rotation rate of $-0.055^{+0.007}_{-0.007}$ kms$^{-1}$pc$^{-1}$ directed at $76^{\circ}$ above the Galactic plane. These maxmimum and minimum rates are asymmetric at the 8$\sigma$ significance level.
    \item We calculated the velocity dispersions in the plane of sky using a Bayesian approach as described in \citet{Armstrong22}. We found a maximum dispersion of $1.225^{+0.048}_{-0.043}$ kms$^{-1}$ directed at $32^{\circ}$ above the Galactic plane and a minimum dispersion of $0.771^{+0.029}_{-0.026}$ kms$^{-1}$ directed at $-62^{\circ}$ below the Galactic plane. These maximum and minimum dispersions are also asymmetric at the 8$\sigma$ significance level.
    \item We performed a 2D linear traceback to estimate the time in the past the $\lambda$ Ori cluster would have been at its minimum area configuration defined as either minimum spanning tree total length \citep{squicciarini21} or the sum of distances between each cluster member and the cluster centre \citep{quintana22}. We found that the cluster would have been at its minimum area configuration $\tau_{\rm min\:area} = 4.1^{+0.1}_{-0.1}$ Myr ago by both definitions.
    \item We identified cluster members consistent with originating from the core radius $r_c$ of the cluster and calculated plane-of-sky traceback times ($\tau_{\rm kin,CA}$) to the point of closest approach of each star to the cluster centre. We compared these times to ages estimated from comparisons to stellar evolution models \citep{baraffe15,marigo17,somers20} $\tau_{iso}$ in a \textit{Gaia} DR3 G-RP versus $M_{G}$ colour-magnitude diagram and found significant spread in $\tau_{kin,CA}$ but little significance in the spread of $\tau_{\rm iso}$.
    \item For cluster members that trace back in the plane of sky to within the cluster core radius of $r_{c} = 0.44$ pc, we found a maximum $\tau_{\rm kin,CA}$ of $4.1^{+0.1}_{-0.1}$ Myr, which is in very close agreement with the time of minimum-area configuration $\tau_{\rm min\: area}$ after discarding \textit{Gaia} DR3 ID 3336145675718723968 as unlikely to have originated from the cluster core on the basis of 3D traceback using an RV measurement from the APOGEE survey. These are candidate cluster members for dynamical ejections from the cluster core.
    \item We found that these candidate cluster core ejectees are distributed with preference to a position angle of $\sim-150^\circ$ (Fig.~\ref{Ejectees}), whereas we would expect stars ejected by either binary supernova or strong dynamical interaction scenarios to be distributed at random position angles. We argue that the recently presented subcluster ejection scenario \citep{polak24} may partially account for the anisotropic distribution we have found. However, it is not expected to produce kinematic age spreads with a range as large as we have observed here. If these candidate ejectees can be confirmed to be consistent with originating from the cluster core in 3D with RV measurements, the significant spread in kinematic ages would indicate a series of dynamical ejections from the cluster core over the $\sim4$ Myr lifetime of the cluster.
    
\end{itemize}

\begin{acknowledgements}
      JA and JCT acknowledge support from ERC Advanced Grant 788829 (MSTAR). This work has made use of data from the ESA space mission \textit{Gaia} (http://www.cosmos.esa.int/gaia), processed by the \textit{Gaia} Data Processing and Analysis Consortium (DPAC, http://www.cosmos.esa.int/web/gaia/dpac/consortium). Funding for DPAC has been provided by national institutions, in particular the institutions participating in the \textit{Gaia} Multilateral Agreement. This research made use of the Simbad and Vizier catalogue access tools (provided by CDS, Strasbourg, France), Astropy \citep{astr13} and TOPCAT \citep{tayl05}.
\end{acknowledgements}

\bibliographystyle{aa} 
\bibliography{aanda}

\end{document}